\documentclass[acmsmall]{acmart}

\renewcommand\footnotetextcopyrightpermission[1]{} 

\pdfoutput=1

\AtBeginDocument{%
  \providecommand\BibTeX{{%
    \normalfont B\kern-0.5em{\scshape i\kern-0.25em b}\kern-0.8em\TeX}}}

\usepackage{tcolorbox}
\usepackage{longtable}

\DeclareUnicodeCharacter{0259}{\v{e}}

\begin{document}

\title{Research on WebAssembly Runtimes: A Survey}

\author{Yixuan Zhang}
\affiliation{%
  \institution{Peking University}
  \city{Beijing}
  \country{China}}
\email{zhangyixuan.6290@pku.edu.cn}

\author{Mugeng Liu}
\affiliation{%
  \institution{Peking University}
  \city{Beijing}
  \country{China}}
\email{lmg@pku.edu.cn}

\author{Haoyu Wang}
\affiliation{%
  \institution{Huazhong University of Science and Technology}
  \city{Wuhan}
  \country{China}}
\email{haoyuwang@hust.edu.cn}

\author{Yun Ma}
\affiliation{%
  \institution{Peking University}
  \city{Beijing}
  \country{China}}
\email{mayun@pku.edu.cn}

\author{Gang Huang}
\affiliation{%
  \institution{Peking University}
  \city{Beijing}
  \country{China}}
\email{hg@pku.edu.cn}

\author{Xuanzhe Liu}
\affiliation{%
  \institution{Peking University}
  \city{Beijing}
  \country{China}}
\email{liuxuanzhe@pku.edu.cn}

\renewcommand{\shortauthors}{Yixuan Zhang et al.}

\begin{abstract}
WebAssembly (abbreviated as Wasm) was initially introduced for the Web but quickly extended its reach into various domains beyond the Web. To create Wasm applications, developers can compile high-level programming languages into Wasm binaries or manually convert equivalent textual formats into Wasm binaries. Regardless of whether it is utilized within or outside the Web, the execution of Wasm binaries is supported by the Wasm runtime. Such a runtime provides a secure, memory-efficient, and sandboxed execution environment designed explicitly for Wasm applications. 
This paper provides a comprehensive survey of research on WebAssembly runtimes. It covers 98 articles on WebAssembly runtimes and characterizes existing studies from two different angles, including the "internal" research of Wasm runtimes(Wasm runtime design, testing, and analysis) and the "external" research(applying Wasm runtimes to various domains). This paper also proposes future research directions about WebAssembly runtimes.
\end{abstract}

\ccsdesc[500]{General and reference~Surveys and overviews}

\keywords{WebAssembly, WebAssembly runtime, WebAssembly System Interface}

\maketitle

\section{Introduction}\label{sec:introduction}
WebAssembly (abbreviated as Wasm) was initially proposed for the Web as a binary instruction format for a stack-based virtual machine. The paper introducing WebAssembly was published in 2017~\cite{bringing}. It outlines the motivation, design, and formal semantics of WebAssembly while offering initial insights into its implementation. Wasm is crafted to serve as a universally compatible compilation target for programming languages, facilitating web deployment for client and server applications~\cite{Wasm-org}. Currently, major web browsers, including Chrome, Firefox, Safari, and Edge, support the execution of WebAssembly (Wasm)~\cite{Wasm-org}. Although Wasm was initially proposed for the Web, it rapidly expanded into multiple domains beyond the Web~\cite{wasm_non_web}, including the Internet of Things (IoT)~\cite{ wen2020wasmachine}, blockchain~\cite{wasm_blockchain, eos-vm, hera}, serverless computing~\cite{wasm_serverless_edge}, edge computing~\cite{mendki2020evaluating, gadepalli2020sledge}, etc.

\begin{figure}[htb]
\centerline{\includegraphics[width=0.75\textwidth]{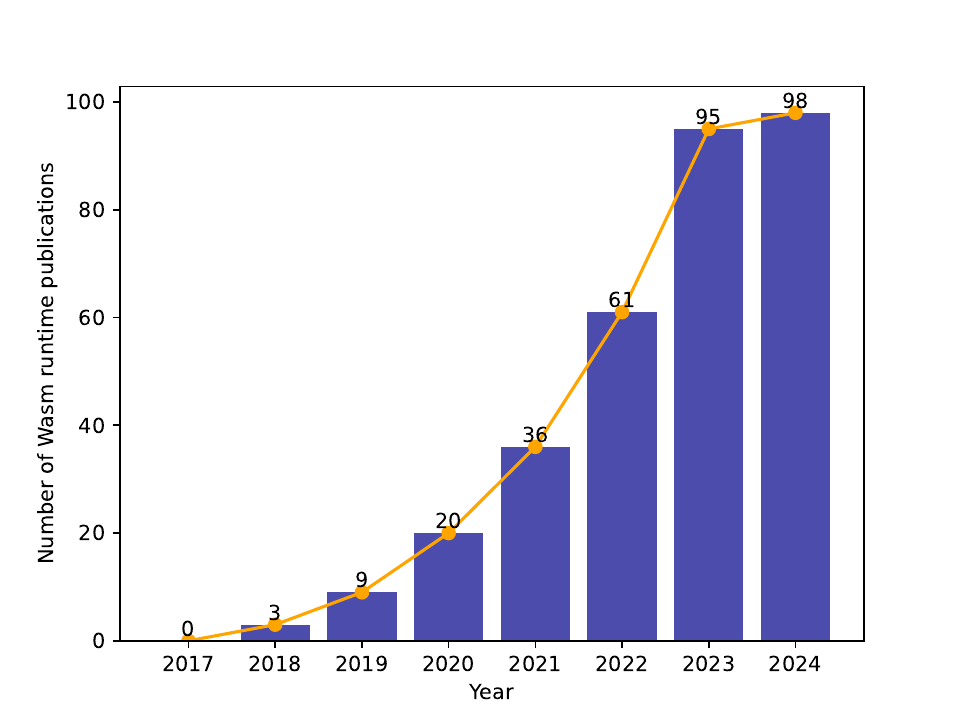}}
	\caption{Wasm runtime publications (accumulative) during 2017-2024. (The publication count for the year 2024 is recorded up to January 8th.)}
	\label{figure:publication}
\end{figure}

To develop Wasm applications, developers can compile high-level programming languages into Wasm binaries or manually convert the equivalent textual format~\cite{watfile} into Wasm binaries. Whether inside or outside the web, the execution of Wasm binaries relies on the Wasm runtime. A Wasm runtime offers a secure, memory-efficient, and sandboxed execution environment tailored for Wasm applications~\cite{Wasm-org}. Specifically, Wasm binaries are executed in the JavaScript engine with JavaScript glue code within the Web~\cite{wang2021empowering}. Outside the Web, Wasm binaries can independently run in standalone Wasm runtimes~\cite{zhang2023characterizing}. This paper will refer to all tools used to execute Wasm binaries as the Wasm runtime. 
Wasm runtime has attracted widespread attention in the academic community since the born Wasm.
Figure~\ref{figure:publication} shows the number of publications on the topic of Wasm runtimes between January 1st, 2017, and January 8th, 2024 (we introduce how we collected these articles in Section ~\ref{sec:article_collection}). The number of papers related to Wasm runtime is rapidly increasing, attracting growing attention to this field.

To the best of our knowledge, this is the first survey of research on Wasm runtimes. We follow the traditional systematic literature review process to collect 98 research papers related to Wasm runtimes. Then we characterize existing studies from two different angles. The first angle is the ``internal'' research of Wasm runtimes, including analysis, testing, and design of Wasm runtimes, aiming to improve the security, performance and scalability of Wasm runtimes. The other angle is the ``external'' research of Wasm runtimes, i.e., applying Wasm runtimes to various kinds of domains (e.g., IoT and edge computing) and addressing the challenges of different scenarios.
Moreover, we discuss the current issues and the future research directions.
As far as we know, this is the first survey focusing on Wasm runtimes. In summary, the paper makes the following contributions:
\begin{itemize}
\item \textit{Definition} This paper defines WebAssembly runtime (Wasm runtime), the workflow of WebAssembly, the architecture and features of Wasm runtimes.
\item \textit{Survey} The paper provides the first comprehensive survey of 98 Wasm runtime articles across various publishing areas such as software engineering, security, and programming languages.
\item \textit{Analyses} This paper illustrates and discusses these related Wasm runtimes in two aspects: application scenarios and research directions.
\item \textit{Future directions} To facilitate the development of Wasm runtimes and the Wasm community, this paper identifies the current issues and promising research directions for Wasm runtimes.
\end{itemize}

\begin{figure}[htb]
\centerline{\includegraphics[width=0.90\textwidth]{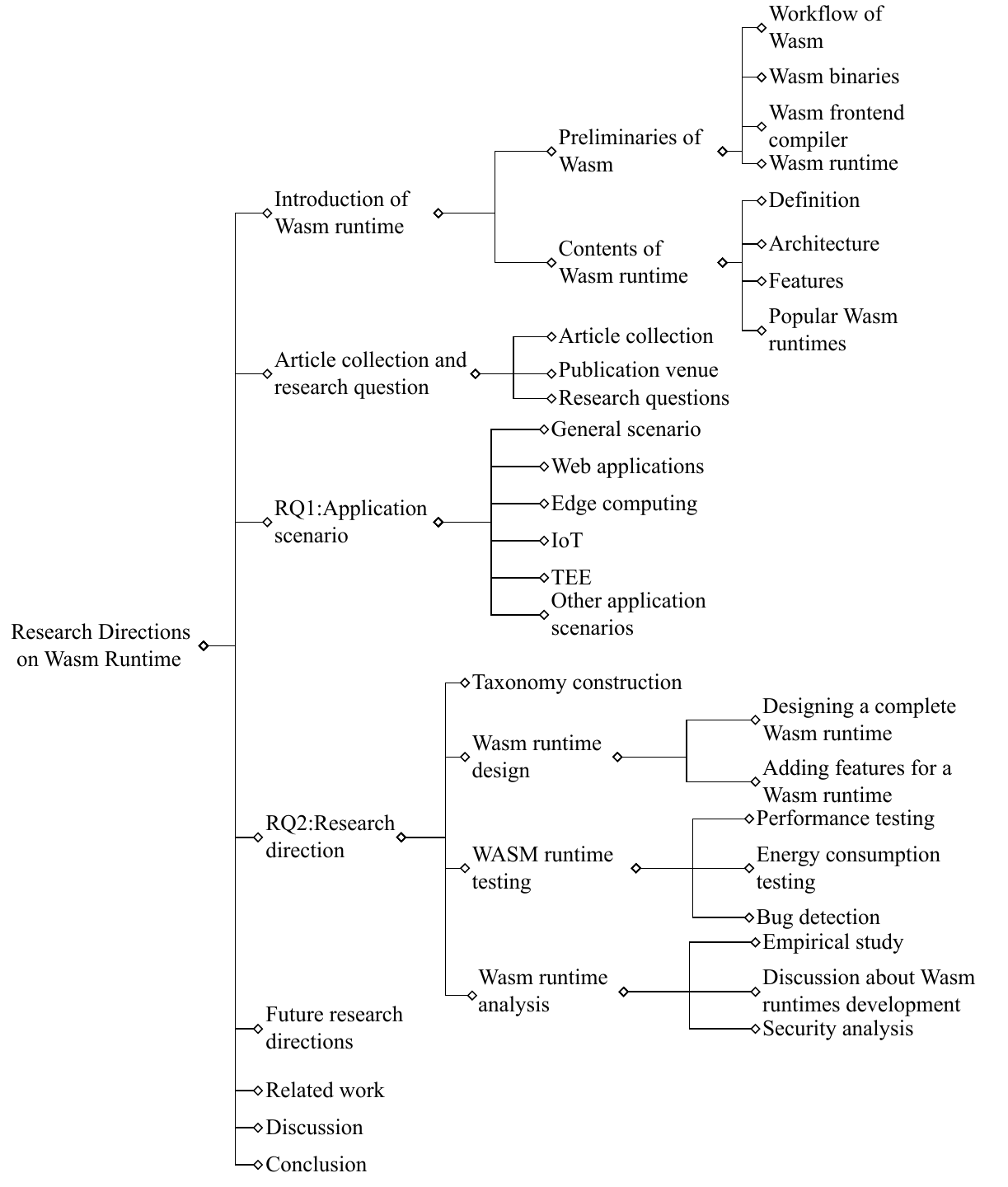}}
	\caption{Tree structure of the contents in this paper.}
	\label{figure:paper_structure}
\end{figure}

Figure~\ref{figure:paper_structure} depicts the paper structure. The details of the review schema can be found in Section~\ref{papercol}.

\section{Preliminaries of WebAssembly}\label{sec:preliminaries}
This section illustrates the background and fundamental terminologies in WebAssembly.
WebAssembly is a binary format that was first proposed for the Web~\cite{bringing} but is now widely used inside and outside the Web, including Internet of things(IoT)~\cite{pham2023webassembly}, edge computing~\cite{gadepalli2020sledge}, and blockchain~\cite{wasm_blockchain}. 

\begin{figure}[htbp]
    \centering
    \begin{minipage}[b]{0.5\textwidth}
        \centering
        \includegraphics[width=\textwidth]{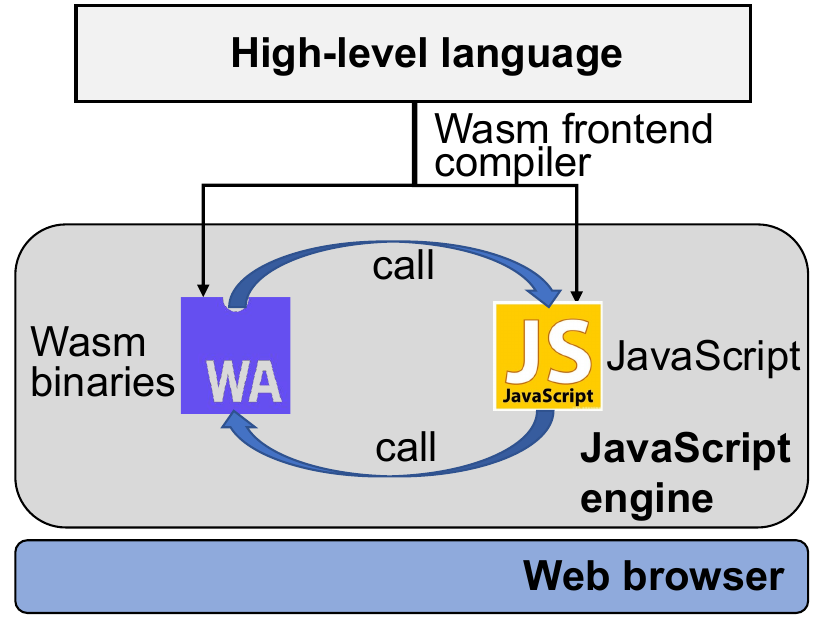}
        \caption{Wasm workflow inside the Web.}
        \label{figure:inside}
    \end{minipage}
    \hspace{0.02\textwidth}
    \begin{minipage}[b]{0.37\textwidth}
        \centering
        \includegraphics[width=\textwidth]{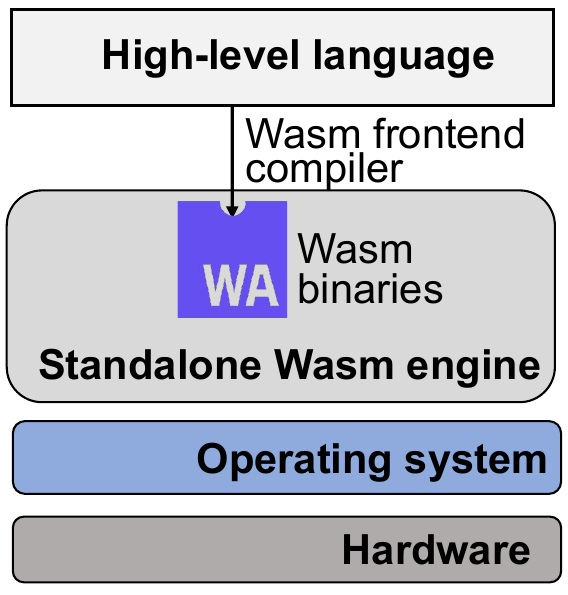}
        \caption{Wasm workflow outside the Web.}
        \label{figure:outside}
    \end{minipage}
\end{figure}

\textbf{Workflow} This section first depicts the workflow of Wasm inside and outside the Web.
As shown in Figure ~\ref{figure:inside}, the Wasm frontend compilers could compile high-level languages, such as C/C++ and Rust, into Wasm binaries with the corresponding JavaScript(JS) glue code. The JS glue code and Wasm binaries are executed in the JS engines in the Web browsers.
As shown in Figure ~\ref{figure:outside}, the Wasm frontend compilers could compile high-level language into Wasm binaries, and the Wasm binaries are executed in standalone Wasm runtimes which are directly deployed in operating systems(OSes).
In this paper, both JS engines that execute Wasm binaries inside the Web and standalone Wasm runtimes that execute Wasm binaries outside the Web are regarded as Wasm runtimes.
That is to say, developers could develop applications inside or outside the Web with high-level languages and generate Wasm binaries to be executed in Wasm runtimes.
The following subsections depict each part of the Wasm workflow.

\begin{figure}[htb]
\centerline{\includegraphics[width=0.90\textwidth]{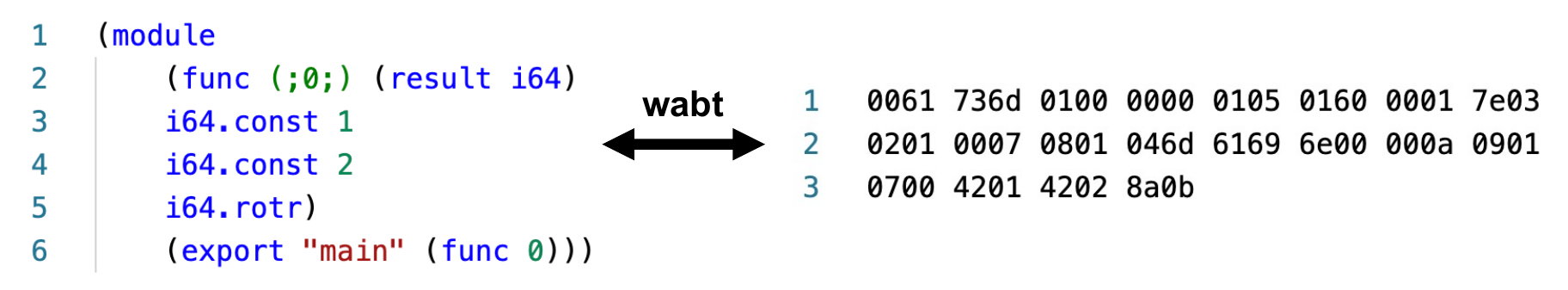}}
	\caption{The conversion between Wasm binaries and textual format.}
	\label{figure:wasm2wat}
\end{figure}

\textbf{Wasm binaries} The WebAssembly (Wasm) binary file is compact, similar to Java class files, and is stored in the file with\texttt{.wasm} suffix~\cite{wasmbook}. 
The WebAssembly (Wasm) specification~\cite{wasmspec} outlines a conceptual stack virtual machine, where the majority of Wasm instructions operate by popping and pushing numbers, leaving the results on the stack.
As shown in Figure ~\ref{figure:wasm2wat}, Wasm binaries could be converted into a pretty-printed textual format (i.e., \texttt{.wat}) with wabt~\cite{wabt} or the reverse process. The wat format~\cite{watfile} is easy to read and can be used to learn the syntax, debug the Wasm code, understand the Wasm modules, manually write Wasm programs, etc.
The Wasm binary format begins with a magic number and a version number, followed by the main body of the Wasm module, as shown in Figure ~\ref{figure:wasm2wat}. A Wasm module is the fundamental unit of Wasm binaries. The specification~\cite{wasmspec} defines the Wasm module in the format of tree nodes. These nodes are written in the style of S-expressions, which goes inside a pair of parentheses \textit{( ... )}. The main body of the Wasm module is organized into different sections, known as segments, where each segment can contain multiple items. These segments include code segments, export segments, function segments, etc.

\textbf{Wasm frontend compiler} Wasm frontend compiler refers to the compilers which could compile high-level language programs into Wasm binaries.
The Wasm frontend compiler first translates high-level language source code into an intermediate representation (IR) and then generates Wasm binaries from the IR~\cite{wasmbook}. During this process, the Wasm frontend compilers include bindings for existing libraries, enabling the use of standard libraries available in the source language within the Wasm runtime~\cite{wasmbook, wasi}. 
As shown in Table~\ref{frontend_compiler},almost all the most widely used high-level languages could be compiled into Wasm binaries through different Wasm frontend compilers.

\begin{table}[t]
\caption{The Wasm frontend compilers.}
\vspace{-0.1in}
\begin{center}
\label{frontend_compiler}
\setlength{\tabcolsep}{1mm}{\begin{tabular}{llrr}
\toprule  
\textbf{Wasm frontend compiler} &\textbf{
Source language} &\textbf{Commits} &\textbf{Stars}\\
\midrule  
Emscripten & C/C++ & 26,521 & 24.8k\\
Rustc & Rust & 243,427 & 88.9k\\
Ppci-Mirror & Python & 1,118 & 0\\
TinyGo & Go & 3,614 & 13.9k\\
Bytecoder & Java & 1,422 & 836\\
AssemblyScript & TypeScript & 1,564 & 16.2k\\
\bottomrule 
\end{tabular}}
\vspace{-0.1in}
\end{center}
\end{table}

\textbf{Wasm runtime} As shown in Figure~\ref{figure:inside} and Figure~\ref{figure:outside}, Wasm binaries could be executed inside and outside the Web with different types of Wasm runtimes. 
Inside the Web, the Wasm runtimes are deployed in the web browsers. Moreover, the Wasm binaries have to be executed with the JS glue code. However, the Wasm runtimes outside the Web are directly deployed on the OSes. And the Wasm binaries could be executed standalonely or interact with high-level languages~\cite{zhang2023characterizing}.

\section{WebAssembly runtime}\label{sec:wasmruntime}
This section gives a definition and analysis of Wasm runtimes. It describes the architecture (components), the most popular Wasm runtimes, and the scenarios of Wasm runtimes.

\subsection{Definition}
The execution of Wasm binaries is defined in an abstract virtual machine (VM). This VM should contain a stack to record the operand values and control constructs and contain an abstract area to store the global state. This VM is expected to support all the instructions in the Wasm binary format and the interaction between Wasm binaries and the environment in which the VM exists~\cite{wasmruntime}.
In this paper, we refer to the term \textit{Wasm runtime} for the programs that could execute Wasm binaries inside or outside the Web.

\subsection{Architecture}
\label{architecture}
\begin{figure}[htb]
\centerline{\includegraphics[width=0.85\textwidth]{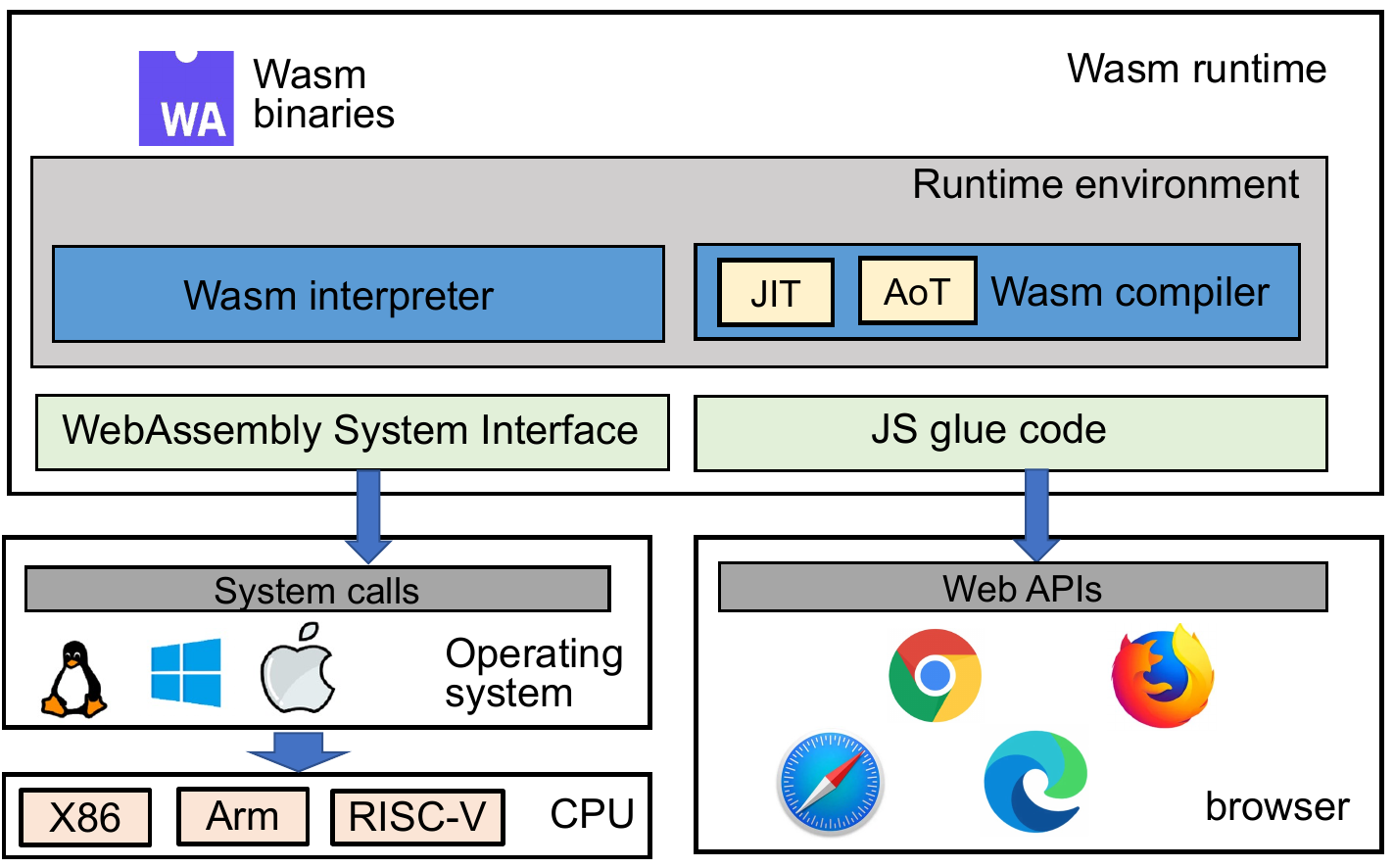}}
	\caption{The architecture of Wasm runtimes.}
	\label{figure:wasmruntime_arch}
\end{figure}

As shown in Figure~\ref{figure:wasmruntime_arch}, a Wasm runtime could be divided into four components~\cite{zhang2023characterizing}, including the Wasm compiler, Wasm interpreter, runtime environment, and the interaction part with the underlying environment (WASI implementation or the JS glue code). 
The execution of Wasm binary instructions includes two ways: interpretation or compilation into native code. Some Wasm runtimes support to execute Wasm binaries with \textbf{Wasm interpreters}, such as wasm-micro-runtime (WAMR)~\cite{WAMR}. Some Wasm runtimes support first compiling the Wasm binaries into an intermediate representation (IR) and then compiling the IR into native code for different hardware, such as wasmtime~\cite{wasmtime} and wasmer~\cite{wasmer}. This paper uses the term \textbf{Wasm compiler} to refer to the compilers that could compile Wasm binaries into native code for different CPUs, including \textit{x86\_64}, \textit{amd64}, etc.
The Wasm runtimes also provide the \textbf{runtime environment} to support allocating memory, performing stack operations, reporting execution error messages, and other features.

In order to \textbf{interact with the environment} the runtime exists, Wasm runtimes in the Web could call the Web APIs through the \textbf{JS glue code}, and standalone Wasm runtimes provide the \textbf{implementation of WebAssembly System Interface (WASI)} to call the system calls. 
WASI is defined by the bytecode alliance~\cite{wasi_preview1}, containing several OS-like features, including the operations of file systems, clock, thread, etc. WASI is the bridge between the sandbox environment and the environment in which it exists. In order to be available in several OSes, WASI unifies the system calls offered by various OSes, such as POSIX in Linux~\cite{wasi_preview1}. 
The standalone Wasm runtimes, such as WAMR~\cite{WAMR}, wasmtime~\cite{wasmtime}, and WasmEdge~\cite{WasmEdge}, could be called in the programs written in high-level languages as a library to allow developers execute Wasm binaries in any possible cases with various languages. And some standalone Wasm runtimes also offer additional tools to users, including Wasm module caching, validation of the Wasm textual file format, and more, such as wasmer.

\subsubsection{Wasm interpreter} 
Wasm is designed as a low-level, cross-platform execution format that can be directly executed by an interpreter. A Wasm interpreter is a tool that interprets WebAssembly binary code instructions individually. One advantage of this approach is that there is no need to wait for a compilation process during loading, which speeds up startup time. However, interpreted execution is generally slower compared to native machine code.
For example, WAMR~\cite{WAMR} and wasm3~\cite{wasm3} provide the interpretation mode for executing Wasm binaries.

\subsubsection{Wasm compiler}
The other way to execute Wasm binaries is to first compile the Wasm bytecode into native code, and then execute the native code. There are two types of Wasm compilers: just in time (JIT) compiler and the ahead of time (AoT) compiler. 
In general, Wasm compilers convert Wasm binaries into their intermediate representation (IR), allocate registers and optimize the IR code to generate the native code. For example, wasmer~\cite{wasmer} use three Wasm compilers, singlepass~\cite{singlepass}, cranelift~\cite{cranelift} and LLVM~\cite{llvm}. 
This process should consider different OSes and CPU architectures. That is to say, the same Wasm binary file could be compiled into different native codes for different OSes and CPUs.

\subsubsection{WebAssembly System Interface}
Although Wasm was first proposed for the Web, it has already extended beyond it. However, when extending Wasm beyond the confines of the browser, a challenge arises. That is how to ensure that Wasm binaries can consistently interact with different OSes, as the Wasm binaries are compiled from different high-level languages. There is where WASI exists~\cite{wasi}.
Specifically, WASI is a middle layer of Wasm runtimes and the underlying OSes. WASI unifies the system calls from various OSes.
The specification of WASI~\cite{wasi_preview1} follows the modular principle, i.e., it allows runtimes to design and implement other modules~\cite{wasidoc} based on the foundational \texttt{wasi-core} module.
The most necessary functions, including files, clocks, and random number generation, are defined in the \texttt{wasi-core} module.
Now, almost all the standalone Wasm runtimes implement the WASI functions, including wasmtime~\cite{wasmtime}, wasmer~\cite{wasmer}, WAMR~\cite{WAMR}, WasmEdge~\cite{WasmEdge}, etc.

\subsection{Features}
\label{feature}
In order to meet the efficient and secure characteristics of Wasm, Wasm runtimes could include the following features~\cite{Wasm-org, li2022bringing}: 
1) \textbf{Safe.} A Wasm runtime is expected to provide a memory-safe, sandboxed environment for executing Wasm binaries, by memory bound checking, etc.
2) \textbf{Efficient.} As Wasm aims to execute at native speed, a Wasm runtime needs to be designed to be efficient and fast.
3) \textbf{Compatible.} Since Wasm is proposed as a general execution binary format, Wasm runtimes needs to be avialble on different platforms.
4) \textbf{Lightweight.} In order to start up at a fast speed and be compatible with resource-constrained platforms, a Wasm runtime needs to be lightweight.

\subsection{Popular Wasm runtimes}
This section illustrates the most popular Wasm runtimes. As shown in Table ~\ref{wasmruntimetable}, according to whether a Wasm runtime could run outside the Web, the Wasm runtimes could be divided into two types: standalone Wasm runtimes and integrated Wasm runtimes.

\begin{table}[t]
\caption{The most popular Wasm runtimes.}
\vspace{-0.1in}
\begin{center}
\label{wasmruntimetable}
\setlength{\tabcolsep}{1mm}{\begin{tabular}{lccccr}
\toprule  
\textbf{Wasm runtime} &\textbf{standalone} &\textbf{interpreter} &\textbf{JIT} &\textbf{AoT} &\textbf{Stars}\\
\midrule  
V8(Chrome/Edge) & $\circ$ & $\circ$ & $\bullet$ & $\circ$ & 22.2k\\
SpiderMonkey(Firefox) & $\circ$ & $\circ$ & $\bullet$ & $\bullet$ & 241\\
WebKit-JavaScriptCore(Safari) & $\circ$ & $\bullet$ & $\circ$ & $\circ$ & 684\\
wasmtime & $\bullet$ & $\circ$ & $\bullet$ & $\bullet$ & 13.8k\\
wasmer & $\bullet$ & $\circ$ & $\bullet$ & $\bullet$ & 16.9k\\
WAMR & $\bullet$ & $\bullet$ & $\bullet$ & $\bullet$ & 4.3k\\
WasmEdge & $\bullet$ & $\circ$ & $\circ$ & $\bullet$ & 7.5k\\
wasm3 & $\bullet$ & $\bullet$ & $\circ$ & $\circ$ & 6.8k\\
\bottomrule 
\multicolumn{5}{l}{$^{\mathrm{*}}$$\bullet$ and $\circ$ refers to the positive and negative value individually.}
\end{tabular}}
\vspace{-0.1in}
\end{center}
\end{table}

\subsubsection{Inside the Web}
When executing Wasm binaries in the Web, Wasm binary files are expected to leverage Web APIs through JS code while supporting the security model in the Web. 
As shown in Table~\ref{wasmruntimetable}, the most widely used four browsers, including Chrome, firefox, safari, and Edge, all support executing Wasm binaries in their JavaScript engines.
\texttt{V8} is a JavaScript and Wasm runtime written in C++, used in Chrome, Edge browser and Node.js~\cite{v8}.
\texttt{SpiderMonkey} serves as Mozilla's engine for JavaScript and WebAssembly, utilized in Firefox, Servo, and several other projects~\cite{spidermonkey}. The implementation is done using C++, Rust, and JavaScript. 
\texttt{WebKit-JavaScriptCore} is the Apple's JavaScript engine, and mainly served for iOS and macOS~\cite{javascriptcore}.
The Wasm runtimes mentioned above are the runtimes that support executing Wasm binaries on the client side of the Web. Moreover, on the Web's server side, Wasm binaries are also supported. For example, Wasm binaries could be executed in Node.js on the server side~\cite{nodejs}.

\subsubsection{Outside the Web}
As shown in Table~\ref{wasmruntimetable}, we illustrate the five most popular Wasm runtimes in GitHub with the top stars. 
\texttt{Wasmtime} is a standalone Wasm runtime proposed by the Bytecode Alliance, designed for both WebAssembly and the WebAssembly System Interface (WASI)~\cite{wasmtime}. It executes WebAssembly code outside the Web, serving as a command-line utility and a library embedded in large-scale applications. It is a highly configurable and embeddable runtime that runs on applications of any scale.
\texttt{Wasmer} is a fast and secure WebAssembly runtime that allows lightweight containers to run anywhere: from desktop applications to the cloud, edge, and IoT devices~\cite{wasmer}. \texttt{Wasmer} supports different compiler frameworks(LLVM~\cite{llvm}, cranelift~\cite{cranelift}, singlepass~\cite{singlepass}) and can generate native code through these Wasm compilers. Similar to \texttt{wasmtime}, \texttt{wasmer} also provides high-level language APIs for developers. The Wasm bytecode can be embedded into programs developed in other languages through these high-level language APIs.
\texttt{WAMR} is a lightweight standalone Wasm runtime with high performance and highly configurable features~\cite{WAMR}, suitable for applications in various cases, including embedded systems, IoT, trusted execution environments(TEE), smart contracts, cloud-native, etc. \texttt{WAMR} supports all the execution modes: interpreter, JIT, and AoT.
\texttt{WasmEdge} is a lightweight, high-performance, and extensible Wasm runtime~\cite{WasmEdge}. It is currently used in edge clouds, serverless software as a service(SaaS) APIs, embedded functions, smart contracts, and intelligent devices.
\texttt{Wasm3} is a high-performance Wasm interpreter~\cite{wasm3}. It is designed to be lightweight, focusing on the size of the runtime executable file and memory usage. This makes it suitable for embedded systems, IoT devices, and other resource-constrained environments.

\section{Article collection and Research question}\label{sec:papercollect}
\label{papercol}
A systematic literature review assesses and interprets all existing research pertinent to a specific subject area. Adhering to Kitchenham's established guidelines~\cite{keele2007guidelines}, we obey the subsequent review protocol in this section.
This section outlines the survey scope, research questions, the methodology used for article collection, an initial analysis of the gathered articles, and the structure of our survey.

\textbf{Survey Scope}
The scope of our article is the software used to execute Wasm binaries, whether on the Web or standalone. When gathering articles, we adhere to the subsequent inclusion criteria. It will be incorporated if an article meets one or more of the listed criteria.
\begin{itemize}
    \item The articles design a Wasm runtime or add features to an existing Wasm runtime.
    \item The articles discuss or test the bugs, performance or security in a Wasm runtime.
    \item The article illustrates the application or future direction of Wasm runtime or analysis of the Wasm runtime.
\end{itemize}

\subsection{Article Collection}
\label{sec:article_collection}
\begin{figure}[htb]
\centerline{\includegraphics[width=0.75\textwidth]{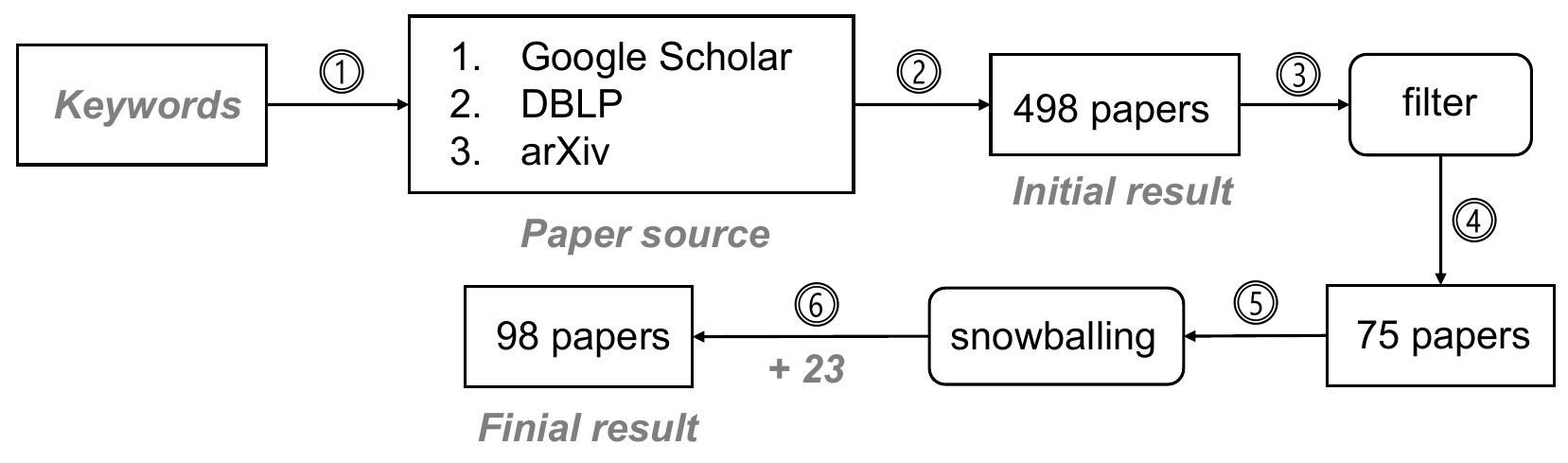}}
	\caption{The article collection process of our methodology.}
	\label{figure:paper_collection}
\end{figure}

We collect relevant published articles about Wasm runtimes to answer the above research questions. This section describes the article collection criteria.
We first explore relevant articles by specifying keywords on widely used search engines and databases. Subsequently, we establish corresponding selection criteria to gather the conclusive articles for the literature review analysis. Figure~\ref{figure:paper_collection} illustrates the detailed process of our methodology for article collection.

\subsubsection{Article Sources.} Following the Zhang.etc's work~\cite{zhang2020machine}, we have chosen three popular scientific databases as article sources for the survey. These sources encompass (1) Google Scholar~\cite{google_scholar}, (2) DBLP~\cite{DBLP}, (3) arXiv~\cite{arXiv}.

\subsubsection{Keywords Searching.} To collect the articles related to Wasm runtimes, we follow the previous work~\cite{hassan2021survey, lo2021systematic, wen2023rise} to define search keywords and apply them to article sources.
First, we include all equivalent names for the Wasm runtime, including runtime, virtual machine (VM), and engine. According to the Wasm runtime architecture~\cite{zhang2023characterizing}, we include the terms to describe components of a Wasm runtime, including compiler, interpreter, and WebAssembly System Interface (WASI).
Second, we consider the terms of Wasm runtime inside Web and outside the Web. We include the names of JS engines to execute Wasm binaries and names of Web browsers, including V8, SpiderMonkey, JavaScriptCore, chrome, firefox, safari, Edge, and browser. We include the names of the most popular Wasm runtimes:\texttt{wasmtime}~\cite{wasmtime}, \texttt{wasmer}~\cite{wasmer}, \texttt{WasmEdge}~\cite{WasmEdge}, \texttt{wasm3}~\cite{wasm3}, and \texttt{WAMR}~\cite{WAMR}.
Third, we also consider features and popular scenarios of Wasm runtime as the keywords, including security, bug, sandbox, trust, performance, IoT, serverless, blockchain, application, Web, embedded system, cloud, and mobile.
Finally, we conducted (2*28 + 1 + 5)*3=186 searches across the three repositories before January 8th, 2024.
\begin{itemize}
    \item Wasm|WebAssembly runtime|virtual machine|engine|VM
    \item Wasm|WebAssembly compiler|interpreter|WASI
    \item WebAssembly System Interface
    \item Wasm|WebAssembly V8|SpiderMonkey|JavaScriptCore
    \item Wasm|WebAssembly chrome|firefox|safari|Edge|browser
    \item wasmtime|wasmer|WasmEdge|WAMR|wasm3
    \item Wasm|WebAssembly performance|security|bug|sandbox|trust
    \item Wasm|WebAssembly IoT|serverless|blockchain|application|Web|embedded system|cloud|mobile
\end{itemize}

Specifically, we collect articles published between 2017 (the publication time of the paper introducing Wasm) and January 8, 2024. Moreover, we obtained 498 articles initially. To find articles that study Wasm runtimes, we only searched for articles whose titles contain these keywords.

\subsubsection{Filtering.} Although some articles may contain relevant keywords, they may be unavailable, repetitive, or not necessarily about Wasm runtime research. For example, they might be about the compilers that compile high-level languages, such as C in Wasm binaries, rather than the component in a Wasm runtime. Therefore, we have established filtering rules to exclude articles from the initial results. We illustrate the filter criteria process to select the relevant research articles about Wasm runtimes.
An article is removed if it satisfies any of the following filter criteria: (1)The complete articles are not available; (2)It is not a research article, such as bachelor, master, or doctoral dissertations; (3)Books, blogs about Wasm runtimes; (4)Moreover, we removed duplicates of articles that appeared in different article sources or searched by different keywords; (5)It is not related to Wasm runtimes. Specifically, we check the remaining articles individually to select the articles related to Wasm runtime. To minimize the risk of erroneous exclusions, the first two authors independently reviewed the articles and discussed whether they were related to Wasm runtime. 
After the filter process, we obtained 75 relevant research articles.

\subsubsection{Snowballing.} Following the commonly adapted snowballing process used in other surveys~\cite{hassan2021survey, lin2022opinion, wen2023rise}, we apply this approach to each obtained article to increase the number of related articles. The snowballing approach includes a forward step and a backward step. These two steps refer to searching the references in each obtained article and other relevant articles from those that cite the obtained articles.
We complement 23 articles to the list. Finally, 98 papers are included in the survey.

\subsection{Publication Venues}
To ascertain the publication venues of research articles on Wasm runtimes, we conduct searches for each article on Google Scholar to identify where they are published, such as conferences or journals.

\begin{figure}[htb]
\centerline{\includegraphics[width=0.95\textwidth]{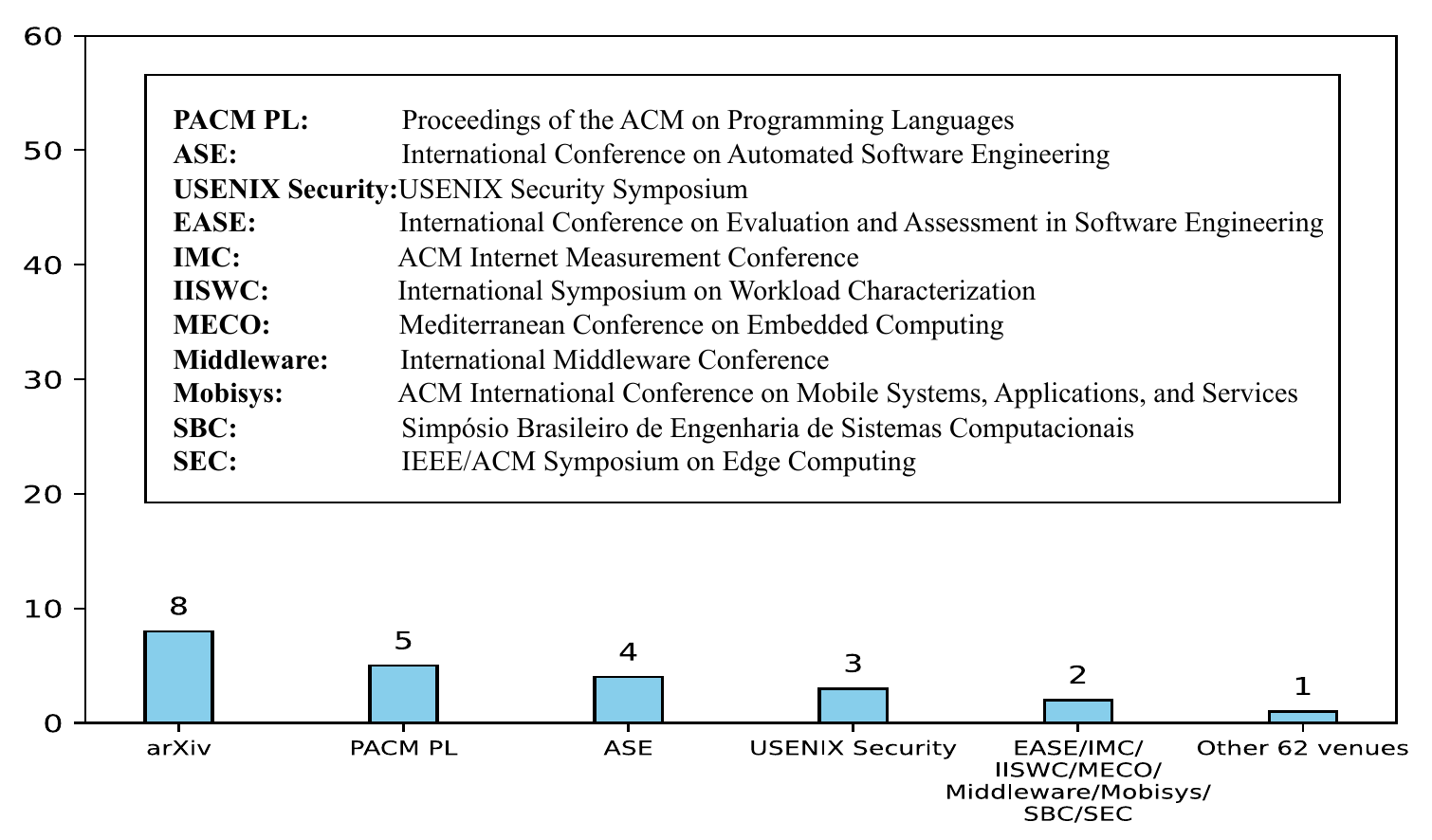}}
	\caption{The main distribution of publication venues for research articles about Wasm runtimes.}
	\label{figure:pub_venue}
\end{figure}

As shown in Figure~\ref{figure:pub_venue}, we examine the publication venues of research articles utilizing the widely recognized Computer Science Rankings~\cite{csranking}. We also explore other significant publication outlets, such as those related to software engineering, database research, and more. Figure~\ref{figure:pub_venue} also presents the publication venues' abbreviation names and corresponding full names. 
Research articles related to Wasm runtime have been published in 74 different venues. Among them, arXiv has the highest proportion, with 8.2\% (8/98) of articles published in arXiv. What's more, 12.2\% (12/98) of articles have been published in PACM PL, ASE, or USENIX Security. Additionally, 16 articles have been published in 8 venues, including IMC, Middleware, etc. Furthermore, the remaining 62 articles are published in different venues, such as PPoPP (ACM SIGPLAN Symposium on Principles \& Practice of Parallel Programming), TOSEM (ACM Transactions on Software Engineering and Methodology), USENIX ATC (USENIX Annual Technical Conference), demonstrating the richness of Wasm runtime-related research across publication venues.

\begin{figure}[htb]
\centerline{\includegraphics[width=0.75\textwidth]{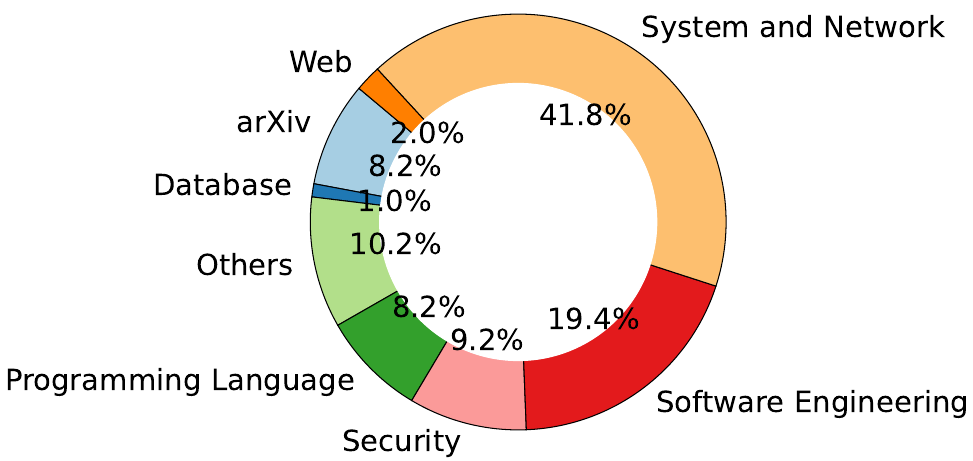}}
	\caption{The publication venue distribution.}
	\label{figure:papersource}
\end{figure}

Figure~\ref{figure:papersource} shows the distribution of papers published in different academic conferences or journals. Among all the papers, 41.8 percent papers are published in conferences and journals about systems and networks, including MobiSys (ACM International Conference on Mobile Systems, Applications, and Services) and INFOCOM (IEEE International Conference on Computer Communications); 19.4 percent of papers are published in software engineering venues, including ASE (International Conference on Automated Software Engineering), SANER (IEEE International Conference on Software Analysis, Evolution, and Reengineering), ICSE(International Conference on Software Engineering), and TOSEM (ACM Transactions on Software Engineering and Methodology). Besides, 9.2\% of the articles are published in conferences and journals about security, such as USENIX Security (USENIX Security Symposium) and CCS (ACM Conference on Computer and Communications Security).
There are small-part papers published in programming languages, databases, the Web, or other venues.
Additionally, 8.2 percent of the papers have not been published in any conference or journal (the arXiv part).

\subsection{Research Questions}
\label{sec:RQ}
To analyze various research directions related to the Wasm runtime, we propose two research questions from two angles.

\begin{itemize}
    \item[RQ1] \textbf{Application Scenario. (the "external" reseach)}\\
    What scenarios do the Wasm runtimes apply in? This research question explores the wide range of scenarios that Wasm could be used for and the role Wasm runtimes play. 
    \item [RQ2] \textbf{Research Direction. (the "internal" reseach)} \\
    What research directions do these articles focus on? This research question delves into the diverse avenues pursued by articles concerning Wasm runtimes. It comprises three sub-questions, delineated in Section~\ref{sec:taxonomy}.
\end{itemize}

\section{RQ1:Application scenario}\label{sec:RQ2}
\begin{figure}[htb]
\centerline{\includegraphics[width=0.85\textwidth]{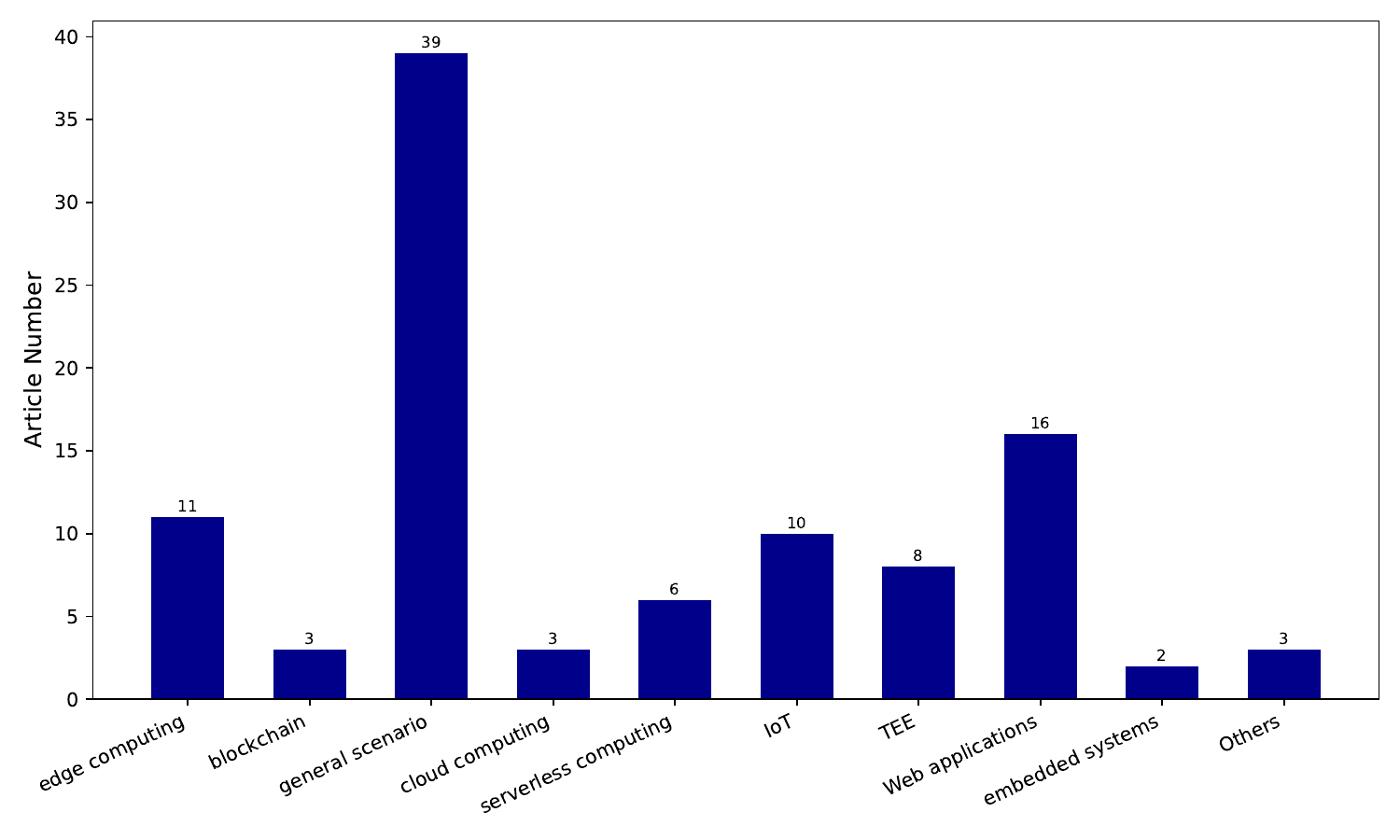}}
	\caption{The application scenarios of the articles.}
	\label{figure:application}
\end{figure}

Although Wasm was initially proposed for the Web, it has been extended to several scenarios.
To answer the research question of which scenarios the existing studies apply Wasm runtimes in, the first authors read all the collected articles and classify the articles according to the application scenarios. This section elaborates on the application scenarios in which the Wasm runtimes are applied.
Some articles apply Wasm runtimes in more than one scenario, so we calculated the data for each scenario when analyzing such papers.
As shown in Figure~\ref{figure:application}, among the articles about Wasm runtimes, the most attention is focused on the general scenario, such as studying the performance of Wasm runtimes that can be universally applied to various scenarios. A significant number of articles also focus on the applications of Wasm runtimes in Web applications, edge computing, and IoT. For example, discussions on the potential and challenges of Wasm runtime in edge computing are prevalent. Some articles also focus on the scenarios of TEE, serverless computing, etc. Five articles focus on other scenarios, including databases, aviation systems, etc. 
The articles spanning multiple application scenarios are counted in all the scenarios to which they belong.
These show the broad adoption of Wasm and Wasm runtimes. This section dived deep into these scenarios.

\subsection{General scenario}
\begin{figure}[htb]
\centerline{\includegraphics[width=0.75\textwidth]{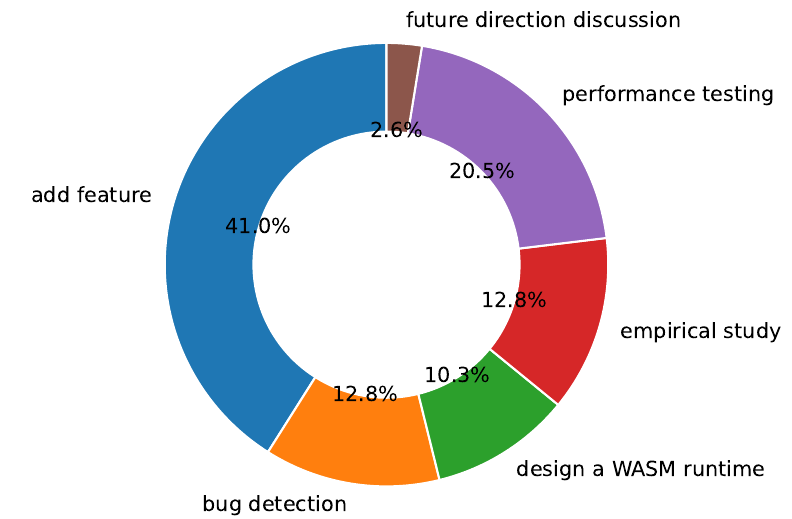}}
	\caption{The research purpose of the articles is for general scenario Wasm runtimes.}
	\label{figure:general_scenario}
\end{figure}

Most articles do not confine Wasm runtime to a specific scenario but instead explore Wasm runtime in a general, universal context.
As shown in Figure~\ref{figure:general_scenario}, more than half of the articles design a new Wasm runtime or add features for existing Wasm runtimes in general scenarios. The other part focuses on bug testing, performance testing, etc.

10.3\% of the articles of general scenarios design a Wasm runtime. For example, 
Bosamiya et al. design a Wasm runtime with the highest priority goal of security guarantee. This Wasm runtime adheres to two principles: employing conventional formal methods to generate mathematically verified safety proofs and integrating Wasm semantics into secure Rust code to generate executable code safely via the Rust compiler~\cite{bosamiya2022provably}.

41.0\% of the articles in general scenarios add features for the Wasm runtime. For example, Titzer et al. design and implement a fast in-place interpreter for Wasm, which could achieve faster startup and lower memory consumption compared to the previous Wasm runtimes~\cite{titzer2022fast}. Zhang et al. focus on linear memory protection in the Wasm runtime~\cite{zhang2023vmcanary}. They first apply a canary approach in code and define two unique Wasm instructions with instrumentation in the Wasm binaries.
Gurdeep Singh et al. broaden the Wasm VM to explore the viability of implementing Wasm on Arduino-compatible microcontrollers~\cite{gurdeep2019warduino}. Specifically, they support 1)safe live code updates, 2)remote debugging at the VM level, and 3)programmer configurable modules to keep the Wasm runtime's footprint, enabling the Wasm runtime to have better performance than the interpreter and increasing the ease of development.

What's more, 20.5\% of the articles of general scenarios test the performance of different Wasm runtimes. For example, Hockley et al. use micro-benchmarks and a macro-benchmark to compare the execution times in Wasm runtimes of Wasm binaries and the native code directly on the machine~\cite{hockley2022benchmarking}. Specifically, they use \texttt{wasmer}~\cite{wasmer} as the testing target, a standalone Wasm runtime that could be used in general scenarios. Ultimately, they discover a performance loss of 5-10 times in the Wasm runtime when compared to native execution.
Szewczyk et al. focus on the performance consumption due to the Wasm's bounds checked memory access safety mechianism~\cite{szewczyk2022leaps}. They extend four popular general scenarios of Wasm runtimes with modern bounds-checking mechanisms, comparing the performance of each with the execution of native code on three different CPU architectures.

There are also 12.8\% articles that provide empirical studies for the general-purpose Wasm runtimes, including characterizing and detecting Wasm runtime bugs~\cite{zhang2023characterizing, wang2023comprehensive}, the issues met by the practitioners of Wasm runtime and application developers~\cite{waseem2023understanding}, characterizing of standalone Wasm runtimes through WABench~\cite{wang2022far}, etc.

Moreover, there are other articles focusing on bug detection, such as WADIFF~\cite{zhou2023wadiff}, WRTester~\cite{cao2023wrtester}, Wasmfuzzer~\cite{jiang2022wasmfuzzer}; some focusing on the discussion of the future directions, such as Joshi's work~\cite{joshianalysis}.

\subsection{Web application}
As shown in Figure~\ref{figure:application}, 16 of 98 articles about Wasm runtimes are in the Web application scenario, which occupies the second position. Wasm was first proposed for the Web, and this field continues to attract widespread research interest.

There are 62.5\%(10/16) papers focusing on performance testing of Wasm execution on the Web. 
For example, Jangda et al. expand BROWSIX into BROWSIX-Wasm, facilitating the direct execution of unmodified Unix applications compiled to Wasm within the browser environment~\cite{jangda2019not}. Moreover, they conduct the performance evaluation of Wasm vs. native through BROWSIX-Wasm. Because of inadequate optimizations and issues with code generation, the performance of Wasm code is reduced by approximately 50\% in Firefox and 89\% in Chrome, on average.
Liu et al. find that while Wasm has the potential to enhance Web XR performance, a performance gap exists between Web-based XR and standalone XR environments.~\cite{liu2023demystifying}.
Through comparison between Wasm and JS in different browsers with the three types of benchmarks, Stotoglou et al. figure out which browser is suitable for which class of problems~\cite{stotoglou2023comparative}.

Besides, three articles dive deep into the energy consumption testing of Wasm on the Web, showing its consumption conservation potential.
Pockstaller et al. find that the execution of Wasm could bring 20\% to 30\% consumption save than JS~\cite{pockstaller2023comparing}. Moreover, the difference in consumption cost is also found between Firefox and Chrome. This study could help reduce carbon dioxide emissions.
Hasselt et al.~\cite {van2022comparing} and Macedo et al.~\cite{de2022webassembly} also find that Wasm could reduce the consumption cost on the Web and improve the battery life of Android devices.

Furthermore, there are 25.0\%(4/16) of the articles about Web applications complement features for the Wasm runtimes~\cite{szanto2018taint, watt2019ct, niessen2020insights, kolosick2022isolation}.

\subsection{Edge computing}
Edge computing occupies a small portion of the application scenarios(11.2\%, 11/98) for Wasm runtimes.
In this scenario, eight runtimes are fully designed or designed based on the foundation of another runtime. 
Sebrechts's runtime~\cite{sebrechts2022adapting} is based on \texttt{wasmtime}~\cite{wasmtime}. They extend the WASI component in \texttt{wasmtime}, enabling Kubernetes controllers being executed as lightweight Wasm modules. The idle controllers will be swap into the disk and activated once they are needed. Finally, the evaluation shows that this runtime achieved a 64\% reduction in memory usage.
To satisfy the essential low latency requirement and the need for lightweight migration on edge services, Nieke et al. design a platform based on Wasm to support portable, secure, and provider-independent services on the edge~\cite{nieke2021edgedancer}.
Besides, different communication stacks also attract attention. The various communication stacks block the reuse of the services across edge computing systems. To tackle this issue, Kreutzer et al. propose a Wasm runtime to run such services as Wasm modules, which oversees the communication between the Wasm module and other components of the system~\cite{kreutzer2023work}.

Two groups of researchers focus on and discuss the future development direction of Wasm in edge computing scenarios.
Hoque et al. explore the potential of utilizing Wasm for edge computing with a comprehensive view. They offer four possible approaches for future development for using Wasm to achieve migratability in edge computing, including code migration, interpreter-based, Wasm-instrumentation, and binary instrumentation~\cite{hoque2022webassembly}.
Kakati et al. offer a systematic overview of the current state of Wasm used in edge computing, including performance optimization, interoperability, and security~\cite{kakati2023webassembly}. They hold the view that Wasm is an ideal resolution with a compact format and high-performance runtime environment for the tasks in edge computing.

\subsection{Internet of things}
In the application scenario of IoT, seven Wasm runtimes are designed fully or based on other runtimes for IoT devices, including Wasmachine~\cite{wen2020wasmachine}, ThingSpire OS~\cite{li2021thingspire}, 
Aerogel~\cite{liu2021aerogel}, 
Wiprog~\cite{li2021wiprog}, WAIT~\cite{li2022bringing}, WOOD~\cite{castillo2021wood} and VM by Jacobsson et al~\cite{jacobsson2018virtual}.

Wasmachine could securely execute Wasm binaries in IoT and Fog devices, with 11\% faster than Linux execution by ahead-of-time compiling and the zero-cost system calls in kernel mode. Wen et al. implement the OS kernel in Rust for the Wasm's memory safety and utilize the sandboxing features provided by the Wasm runtime~\cite{wen2020wasmachine}.
Li et al. design an IoT operating system that uses AoT compilation to achieve seamless inter-module communication when executing Wasm binaries with levels of optimizations~\cite{li2021thingspire}.
Aerogel strongly emphasizes access control, addressing the disparity between bare-metal IoT devices and Wasm runtimes~\cite{liu2021aerogel}. The Wasm runtime is crafted as a multi-tenant platform, treating each Wasm application as a distinct tenant. The sandbox feature within the Wasm runtime plays a pivotal role in effectively managing access control for sensors or actuators~\cite{liu2021aerogel}.
Li et al. provide an integrated programming approach for programmers to easily develop IoT applications in Wiprog~\cite{li2021wiprog}. Programmers could utilize the API offered by Wiprog to access the IoT device operations and assign the placement of the operation by annotations.
WAIT is designed as a secure and energy-efficient WebAssembly (Wasm) runtime. Similarly, to enhance execution efficiency, WAIT also employs Ahead-of-Time (AoT) compilation~\cite{li2022bringing}.
Differently, Jacobsson et al. devise a WebAssembly (Wasm) runtime specifically crafted for wearable devices, employing an interpreter-based approach to enable over-the-air programming through Bluetooth low energy~\cite{jacobsson2018virtual}.

In general, due to the specific nature of IoT devices, the primary considerations in the design of Wasm runtimes are performance and energy consumption. In addition to this, researchers have customized distinct Wasm binary execution strategies and interfaces for various devices.

\subsection{Trusted execution environment}
As shown in Table~\ref{TEE_Wasm_runtime}, most of the articles (7/8, 87.5\%) about TEE application for Wasm runtimes focus on designing new WebAssembly (Wasm) runtimes entirely or based on existing Wasm runtimes for Trusted Execution Environment (TEE) environments, and only one article analyzes the security issues for extending Wasm runtimes in TEE.

\begin{table}[t]
\caption{The Wasm runtimes designed for TEEs.}
\vspace{-0.1in}
\begin{center}
\label{TEE_Wasm_runtime}
\begin{tabular}{p{0.2\linewidth}p{0.2\linewidth}p{0.4\linewidth}}
\toprule  
\textbf{Wasm runtime} & \textbf{TEE architecture} & \textbf{Description} \\
\midrule 
TWINE~\cite{menetrey2021twine, menetrey2023comprehensive} & Intel SGX & A secure and trusted version of SQLite within a TEE.\\
The runtime by Ménétrey et al.~\cite{menetrey2024holistic} & Different TEE architectures & A secure and attested pub/sub system.\\
PRIVATON~\cite{subramanyan2023privaton} & Different TEE architectures & A runtime based on finite state automatons. \\
The runtime by Pop et.al~\cite{pop2022towards} & Arm TrustZone & An enclave software design based on Wasm runtime. \\
AccTEE~\cite{goltzsche2019acctee} & Intel SGX & A two-way sandbox offering accounting of resource usage. \\
WATZ~\cite{menetrey2022watz} & Arm TrustZone & A runtime for the efficient and secure execution of Wasm binaries.\\
\bottomrule 
\end{tabular}
\vspace{-0.1in}
\end{center}
\end{table}

Ménétrey et al. first designed a Wasm runtime based on the existing standalone Wasm runtime: \texttt{WAMR}~\cite{WAMR} in TEE in 2021~\cite{menetrey2021twine} and evalute the performance of the designed runtime in 2023~\cite{menetrey2023comprehensive}. To implement a secure version of the SQLite, they extend the WASI component of \texttt{WAMR} to interact with the underlying system calls and secure libraries supported by Intel SGX. 
When TWINE was implemented in production with the company Credora, the results demonstrated a performance range from a 0.7x slowdown to a 1.17x speedup compared to the native runtime. The study shows that library optimization could improve the performance to a notable 4.1× speedup.
Furthermore, they propose the first Wasm runtime for Arm TrustZone in 2022~\cite{menetrey2022watz}. WATZ is a trusted Wasm runtime with remote attestation capabilities, suitable for TrustZone and applicable to IoT.
Moreover, they extended Wasm runtimes to various TEE architectures in 2023~\cite{menetrey2024holistic}. They designed a sub/pub system in 2023, modifying the TLS protocol to introduce mutual authentication in network communications.

As for privacy preservation, TEE is a good choice, and some researchers combine TEE and Wasm runtimes for privacy preservation. Subramanyan et al. design a Wasm runtime for the general TEE architecture, utilizing the potential of TEE and Wasm to privacy-preserving computations~\cite{subramanyan2023privaton}. PRIVATION lets users define a set of privileges under control, enabling explicit control over computation capabilities.
Pop et al. extend the existing Wasm runtime \texttt{Wasmi}~\cite{wasmi} to run services in TEE to protect the privacy-sensitive data~\cite{pop2022towards}.

As for the remote computation field, due to the trust gap between code providers and code executors, the computation is always executed in a sandbox. Goltzsche et al. provide a two-way sandbox Wasm runtime to account for the resource usage of remote computation~\cite{goltzsche2019acctee}.

\subsection{Other application scenarios}
Wasm runtimes are also applied in other scenarios, such as serverless computing~\cite{kjorveziroski2022evaluating, zhao2023reusable, gackstatter2022pushing, marcelino2023cwasi, kjorveziroski2023webassembly_orchiestration, kjorveziroski2023webassembly}, blockchain~\cite{almstedt2023contractbox, yang2020seraph, zheng2020vm}, embedded systems~\cite{wallentowitz2022potential, scheidl2020valent}, aviation systems~\cite{zaeske2023webassembly}, etc.

Gackstatter et al. design a Wasm runtime, WOW, as a serverless container runtime, with the reduction of cold-start latency by 99.5\%, suitable for various serverless computing workloads~\cite{gackstatter2022pushing}.
Kjorveziroski et al. evaluate the performance of Wasm execution in serverless computing~\cite{kjorveziroski2023webassembly}. They apply three widely used Wasm runtimes: \texttt{wasmtime}~\cite{wasmtime}, \texttt{wasmer}~\cite{wasmer} and \texttt{WasmEdge}~\cite{WasmEdge} to compare the cold start delay and the total execution times, finding that Wasm runtimes perform faster on 10 of 13 tests compared to container runtimes. Moreover, \texttt{wasmtime} is the best.

The blockchain community also embraces Wasm, such as Ethereum, which favors the smart contracts to be executed as Wasm binaries on Ethereum 2.0~\cite{eth2.0}.
Zheng et al. compare the performance of smart contracts executed in Wasm or in the previous language used in Ethereum, Solidity, however, finding that executing smart contracts in Wasm is still far from ideal~\cite{zheng2020vm}. 
Yang et al. propose a tool called Seraph to automatically analyze the security issues for the platforms executing smart contracts in Wasm~\cite{yang2020seraph}.

Moreover, it is promising that Zaeske et al. bring Wasm into aviation systems in 2023~\cite{zaeske2023webassembly}. They integrate the Wasm interpreter into the current avionics software stack, utilizing Wasm to implement the common application binary interface.

In general, Wasm shows significant application value in various fields, and the execution of these Wasm applications relies heavily on the design, updates, and analysis of the Wasm runtimes.

\begin{tcolorbox}
\textbf{Summary of answers to RQ1:} \\
\textbf{(1)} Due to compactness, safety, and other features, Wasm has been widely used in different scenarios inside or outside the Web, including Web applications, edge computing, IoT, blockchain, embedded systems, serverless computing, etc. 
39.8\% (39/98) of the articles do not confine the runtime to a specific scenario, which could be used in a general scenario.\\
\textbf{(2)} Wasm runtimes are fundamental to utilizing Wasm in various scenarios. They support the safe, efficient execution of Wasm binaries. High-quality Wasm runtime promotes broader applications of Wasm, and the widespread use of Wasm also imposes higher requirements on Wasm runtime.
When introducing Wasm into a new scenario, the Wasm runtime needs to be modified to adapt to the new environment, considering the characteristics of the scenario.
\end{tcolorbox}

\section{RQ2:Research direction}\label{sec:RQ3}
\subsection{Taxonomy and sub-questions}
\label{sec:taxonomy}

\subsubsection{Taxonomy construction}
\begin{figure}[htb]
\centerline{\includegraphics[width=0.98\textwidth]{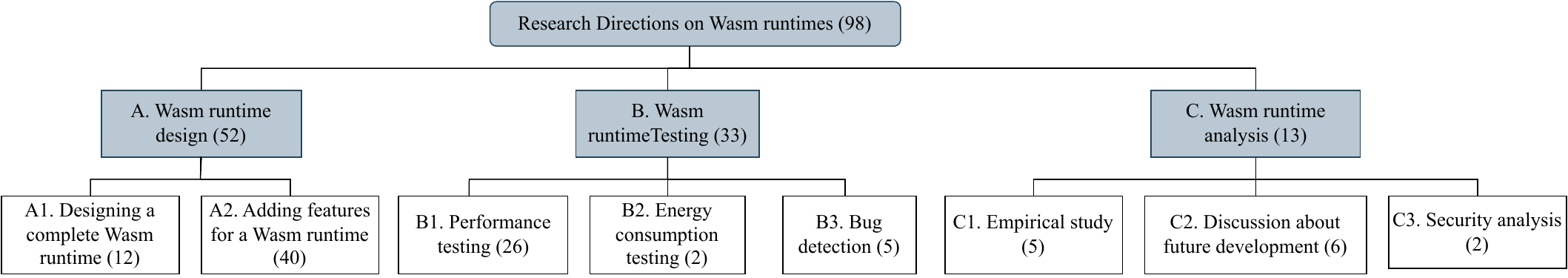}}
	\caption{A taxonomy of research directions in the Wasm runtime literature.}
	\label{figure:research_directions}
\end{figure}

As shown in Figure~\ref{figure:research_directions}, we construct a taxonomy of the research directions of the Wasm runtime literature, following the methodology used in other taxonomy construction works~\cite{zhang2023characterizing, chen2021empirical, wen2023rise}. 
No more than ten articles could focus on more than one direction; as for these articles, we assign them according to the primary purpose. For example, WAFL detects bugs~\cite{hassler2021wafl}. However, the primary purpose is bug detection, with adding features to achieve the goal. Thus, this article is divided into the \textit{bug detection} category.
Our taxonomy includes three root categories of research directions in the grey boxes, such as \textit{Wasm runtime design} and \textit{testing}. Moreover, the number in the parentheses represents the specific data of articles in the corresponding categories. For example, 33 articles test the Wasm runtimes. 
Furthermore, the boxes in the white color represent the sub-directions under a specific research direction scope. For instance, the category \textit{B. Wasm runtime testing} contains three sub-categories, including \textit{preformance testing}, \textit{energy consumption testing} and \textit{bug detection}. In total, there are 33 articles focusing on the testing of Wasm runtimes.

This sub-section will explain our taxonomy in detail. 
First, articles focusing on the design of Wasm runtimes account for 53.1\% (52/98) of all the articles about Wasm runtimes, which occupies the most significant portion. This direction contains 23.1\% (12/52) design of a complete Wasm runtime for different usage (\textit{A1. Designing a complete Wasm runtime}), 76.9\% (40/52) complement features based on fundamental Wasm runtimes (\textit{A2. Adding features for a Wasm runtime}).

Second, in the taxonomy of Wasm runtime literature, \textit{testing} takes the second portion of the articles. In this category, researchers mainly focus on \textit{performance testing} of Wasm execution, which accounts for 78.8\% (26/33). The high performance of Wasm is a significant highlight, and as a result, performance issues have captured enough interest among researchers. Furthermore, there are 6.1\% (2/33) and 15.2\% (5/33) articles that focus on the \textit{energy consumption} and \textit{bug detection} individually.

Third, our taxonomy shows that the analysis of Wasm runtimes accounts for the third largest percentage, i.e., 13.3\% (13/98). About half of these studies (5/13) are empirical studies about Wasm runtimes. Furthermore, there are also 46.2\% (6/13) articles discuss the future development directions, and 15.4\% (2/13) analyze the security issues in Wasm runtimes.

\begin{table}[t]
\caption{The Publication Venues and their Covered Research Directions.}
\vspace{-0.1in}
\label{pub_tab}
\setlength{\tabcolsep}{1mm}
\begin{tabular}{p{0.3\linewidth} p{0.7\linewidth}}
\toprule  
\textbf{Publication venues} &\textbf{Covered research directions} \\
\midrule  
arXiv & Adding features for a Wasm runtime, Bug detection, Empirical study, Security analysis, Performance testing\\
\cline{2-2}
PACM PL & Adding features for a Wasm runtime\\
\cline{2-2}
ASE & Performance testing, Bug detection\\
\cline{2-2}
USENIX Security & Adding features for a Wasm runtime, Designing a complete Wasm runtime\\
\cline{2-2}
EASE/IMC/IISWC/MECO/\newline
Middleware/Mobisys/\newline
SBC/SEC & Performance testing, Energy consumption testing, Empirical study, Adding features for a Wasm runtime, Discussion about future directions, Designing a complete Wasm runtime\\
\cline{2-2}
Other 62 venues & All the leaf research directions.\\
\bottomrule 
\multicolumn{2}{l}{$^{\mathrm{*}}$The publication venues are in abbreviation version.}
\end{tabular}
\vspace{-0.1in}
\end{table}

As shown in Table~\ref{pub_tab}, we summarize the main publication venues and the corresponding research directions. The articles in arXiv cover 62.5\%(5/8) leaf categories, focusing on the diverse research directions of Wasm runtimes. The articles published in ASE mainly focus on the Wasm runtime testing, aligned with the scope of software engineering conferences. The articles published in other 62 venues cover all the leaf research directions.

\subsubsection{Sub-questions}

According to the constructed taxonomy, we propose four sub-questions for RQ2:
\begin{itemize}
    \item[RQ2-1] \textbf{Wasm runtime design.} What components are modified or considered in designing a Wasm runtime? This sub-research question investigates the components and features (what to consider and how to design) in the articles designing Wasm runtimes.
    \item[RQ2-2] \textbf{Wasm runtime testing.} What benchmarks and methods are used in testing Wasm runtimes? This sub-research question investigates the benchmarks (how to build and the benchmark content) and methods (how to test and what to test) used in the articles testing Wasm runtimes.
    \item[RQ2-3] \textbf{Wasm runtime analysis.} What aspects do the empirical studies focus on? What directions do they propose for Wasm runtimes? What are the security issues they are concerned about?
\end{itemize}

\subsection{RQ2-1:Wasm runtime design}

Most researchers in the Wasm runtime field seek to contribute to the Wasm runtime itself. 23.1\% (12/52) of researchers contribute to the overall runtime design, and 76.9\% (40/52) introduce new features to the runtime.
This section compares the Wasm runtime architectures and modified Wasm runtime components of the 52 related articles.

\begin{figure}[htbp]
    \centering
    \begin{minipage}[b]{0.45\textwidth}
        \centering
        \includegraphics[width=\textwidth]{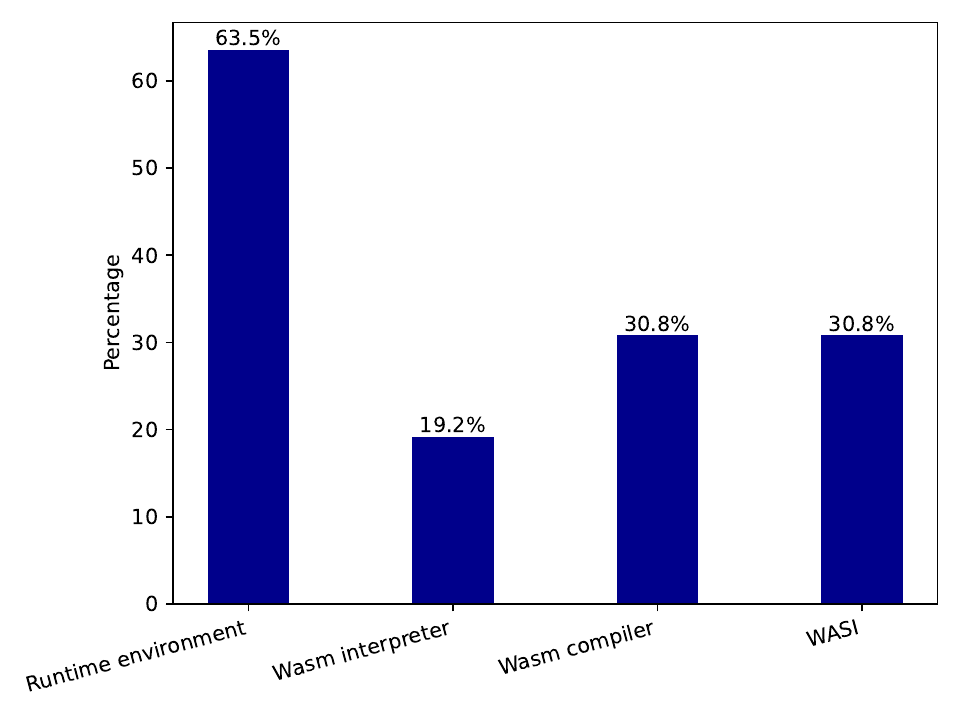}
        \caption{The components modified or mainly considered when designing Wasm runtimes.}
        \label{figure:component}
    \end{minipage}
    \hspace{0.02\textwidth}
    \begin{minipage}[b]{0.45\textwidth}
        \centering
        \includegraphics[width=\textwidth]{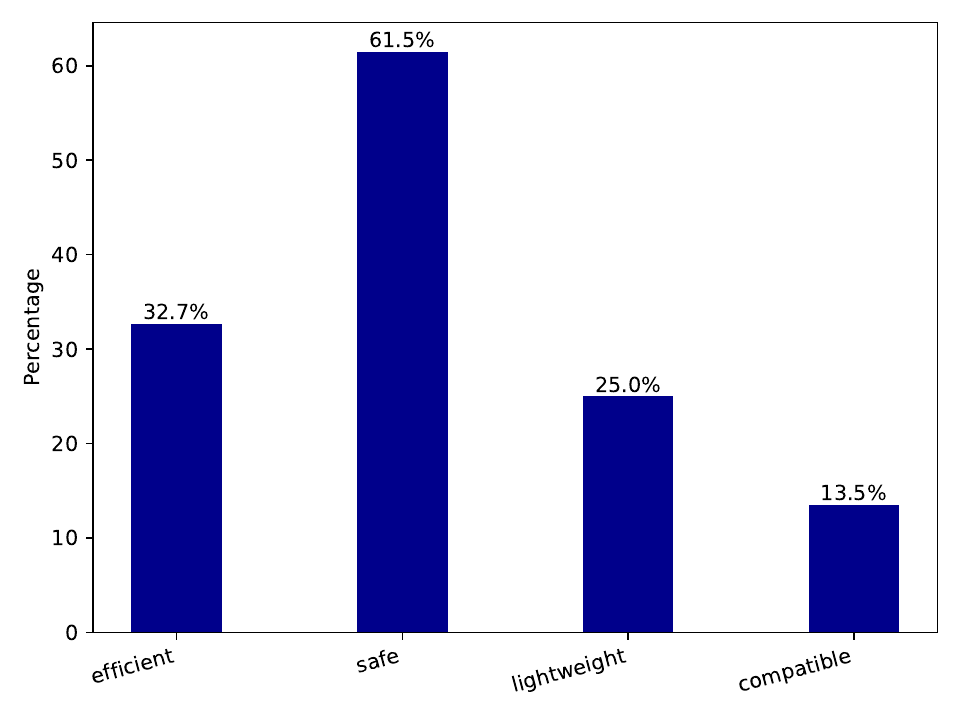}
        \caption{The features utilized when designing Wasm runtimes.}
        \label{figure:feature}
    \end{minipage}
\end{figure}

Figure~\ref{figure:component} illustrates the proportions of various components mainly considered or utilized in designing the Wasm runtime. Due to the consideration of multiple components in the design of some runtimes, these runtimes are included in the counts for different component proportions. 
More than half of the papers in the total 52 Wasm runtime designing articles(\textit{Category A}) focus on the runtime environment, including linear memory protection~\cite{song2023metasafer, lei2023put, zhang2023vmcanary}, safety checking~\cite{geller2024indexed}, workload scheduling~\cite{zhu2022lawow}, etc. 

Figure~\ref{figure:feature} shows the percentages of Wasm runtime features utilized when designing Wasm runtimes. An article could be counted more than once since some leverage multiple features.
The safety feature, encompassing memory safety, sandbox mechanisms, and more, is the most compelling aspect capturing researchers' attention. Over half of the 52 articles (\textit{Category A.1}) primarily utilize or contribute to the safety feature of Wasm runtimes.

As shown in Table~\ref{designed_Wasm_runtime}, the Wasm runtimes designed by the researchers are listed, including the runtimes designed completely and the runtimes designed by adding features. The content \textit{original} in the fourth line means the runtime is designed thoroughly without being based on other runtimes.

\begin{center}
\begin{longtable}{ p{0.2\textwidth} | p{0.28\textwidth} | p{0.25\textwidth} | p{0.22\textwidth}}
\caption{The runtimes designed by the researchers.}
\label{designed_Wasm_runtime} \\
\toprule
\textbf{Wasm runtime} & \textbf{Components} & \textbf{Features} & \textbf{Based Runtime}\\
\midrule
Sledge~\cite{gadepalli2020sledge} & Wasm compiler(AoT), runtime environment & safe, lightweight & original\\
\cline{1-4}
TruffleWasm~\cite{salim2020trufflewasm} & Wasm interpreter, Wasm compiler(JIT), runtime environment, WASI & safe, efficient & original\\
\cline{1-4}
Wasmachine~\cite{wen2020wasmachine} & Wasm compiler(AoT), runtime environment, WASI & safe, efficient & original\\
\cline{1-4}
Bosamiya et al.'s runtime~\cite{bosamiya2022provably} & Wasm compiler(AoT), runtime environment & safe, efficient & original\\
\cline{1-4}
WAIT~\cite{li2022bringing} & Wasm compiler(AoT), runtime environment, WASI & safe, lightweight, efficient & original\\
\cline{1-4}
WATZ~\cite{menetrey2022watz} & Wasm compiler(AoT), runtime envioroment, WASI & safe, efficient & original\\
\cline{1-4}
WaVe~\cite{johnson2023wave} & runtime environment, WASI & safe & original\\
\cline{1-4}
Jacobsson et al.'s runtime~\cite{jacobsson2018virtual} & Wasm interpreter, WASI & lightweight & original\\
\cline{1-4}
eWasm~\cite{peach2020ewasm} & Wasm compiler(AoT), runtime environment & safety, lightweight & original\\
\cline{1-4}
ThingSpire OS~\cite{li2021thingspire} & WASM compiler(AoT) & compatible, safe & original\\
\cline{1-4}
WasmAndriod~\cite{wen2022wasmandroid} & Wasm compiler(AoT), runtime environment & compatible & original\\
\cline{1-4}
PRIVATION~\cite{subramanyan2023privaton} & runtime environment, WASI & safe & original\\
\cline{1-4}

WiProg~\cite{li2021wiprog} & Wasm interpreter, Wasm compiler(JIT), runtime environment, WASI & compatible & wasmer, wasm3\\
\cline{1-4}
AccTee~\cite{goltzsche2019acctee} & runtime envrionment & safe & V8 engine\\
\cline{1-4}
TWINE~\cite{menetrey2021twine, menetrey2023comprehensive} & WASI & lightweight, safe & WAMR\\
\cline{1-4}
Aerogel~\cite{liu2021aerogel} & runtime environment & safe & WAMR\\
\cline{1-4}
Titzer et al.'s runtime~\cite{titzer2022fast} & Wasm interpreter & efficient & Wizard\\
\cline{1-4}
Pop et al.'s runtime~\cite{pop2022towards} & Wasm interpreter & compatible & WASMI\\
\cline{1-4}
ContractBox\cite{kohn2022duckdb} & WASI & safe & WAMR\\
\cline{1-4}
metaSafer~\cite{song2023metasafer} & runtime environment & safe & JS VM\\
\cline{1-4}
WasmRef-Isabelle~\cite{watt2023wasmref} & Wasm interpreter & efficient & wasmtime\\
\cline{1-4}
PKUWA~\cite{lei2023put} & runtime environment & safe & wasmtime\\
\cline{1-4}
Moron et al.'s runtime~\cite{moron2023support} & Wasm compiler(JIT) & efficient & wasm3\\
\cline{1-4}
MPIWasm~\cite{chadha2023exploring} & WASI, runtime envrionment & lightweight, safe, compatible & wasmer\\
\cline{1-4}
Zhao et al.'s runtime~\cite{zhao2023reusable} & runtime environment & safe & WAMR\\
\cline{1-4}
M{\'e}n{\'e}trey et al.'s runtime~\cite{menetrey2024holistic} & runtime environment & safe & TWINE\\
\cline{1-4}
Sebrechts et al.'s runtime~\cite{sebrechts2022adapting} & WASI & lightweight & wasmtime\\
\cline{1-4}
Mäkitalo et al.'s runtime~\cite{makitalo2021bringing} & runtime environment, Wasm compiler(JIT) & efficient & wasmtime\\
\cline{1-4}
WASP~\cite{marques2022concolic} & Wasm interpreter & safe & Wasm interpreter~\cite{wasminterpreter}\\
\cline{1-4}
Gu et al.'s runtime~\cite{gu2023constant} & Wasm compiler(JIT) & safe & wasmtime\\
\cline{1-4}
Watt et al.'s runtime~\cite{watt2019ct} & runtime environment & safe & V8 engine\\
\cline{1-4}
Gobi~\cite{narayan2019gobi} & WASI & safe & wasmtime\\
Wasm-precheck~\cite{geller2024indexed} & runtime environment & safe & wasmtime\\
\cline{1-4}
Nie{\ss}en et al.'s runtime~\cite{niessen2020insights} & Runtime environment & efficient & V8 engine\\
\cline{1-4}
Kolosick et al.'s runtime~\cite{kolosick2022isolation} & Wasm compiler(AoT), runtime environment & safe, efficient & Lucet\\
\cline{1-4}
LOWOW~\cite{zhu2022lawow} & runtime environment & lightweight, efficient, compatible & WasmEdge\\
\cline{1-4}
Nomad~\cite{nurul2021nomad} & Wasm interpreter & compatible & wasm3\\
\cline{1-4}
Abbadini et al.'s runtime~\cite{abbadini2023poster} & WASI & safe & WasmEdge, wasmtime\\
\cline{1-4}
WOW~\cite{gackstatter2022pushing} & runtime environment & lightweight & wasmtime, wasmer, WAMR\\
\cline{1-4}
Swivel~\cite{narayan2021swivel} & Wasm compiler(AoT) & safe & Lucet\\
\cline{1-4}
Szanto et al.'s runtime~\cite{szanto2018taint} & runtime environment & safe & JS VM\\
\cline{1-4}
VMCANARY~\cite{zhang2023vmcanary} & c safe & wasmtime\\
\cline{1-4}
WARDuino~\cite{gurdeep2019warduino} & runtime environment & efficient & The runtime by Joel Martin~\cite{cazzola2016dodging}\\
\cline{1-4}
Wasm/k~\cite{pinckney2020wasm} & Wasm compiler(JIT) & efficient & wasmtime\\
\cline{1-4}
WOOD~\cite{castillo2021wood} & runtime environment & lightweight & WARDuino\\
\cline{1-4}
Zaeske et al.'s runtime~\cite{zaeske2023webassembly} & Wasm interpreter & compatible & Wasmi\\
\cline{1-4}
Kjorveziroski et al.'s runtime~\cite{kjorveziroski2023webassembly_orchiestration} & runtime environment & efficient & wasmtime\\
\cline{1-4}
EDGEDANCER~\cite{nieke2021edgedancer} & runtime environment, WASI & efficient, lightweight & WAMR\\
\cline{1-4}
Nakakaze et al.'s runtime~\cite{nakakaze2022retrofitting} & Wasm interpreter & lightweight & wasm3, V8\\
\cline{1-4}
CWASI~\cite{marcelino2023cwasi} & runtime environment & safe & WasmEdge\\
\cline{1-4}
Kreutzer et al.'s runtime~\cite{kreutzer2023work} & runtime environment & safe & wasmer\\
\bottomrule
\multicolumn{4}{l}{$^{\mathrm{*}}$The second column refers to the components mainly modified in the runtime.}\\
\multicolumn{4}{l}{$^{\mathrm{*}}$The third column refers to the main contributed designed features.}
\end{longtable}
\end{center}

\subsubsection{Designing a complete Wasm runtime}

To study the design of a complete Wasm runtime and what features of the Wasm runtime are primarily considered or utilized, this section is organized by the features mainly focused on.

\begin{figure}[htb]
\centerline{\includegraphics[width=0.7\textwidth]{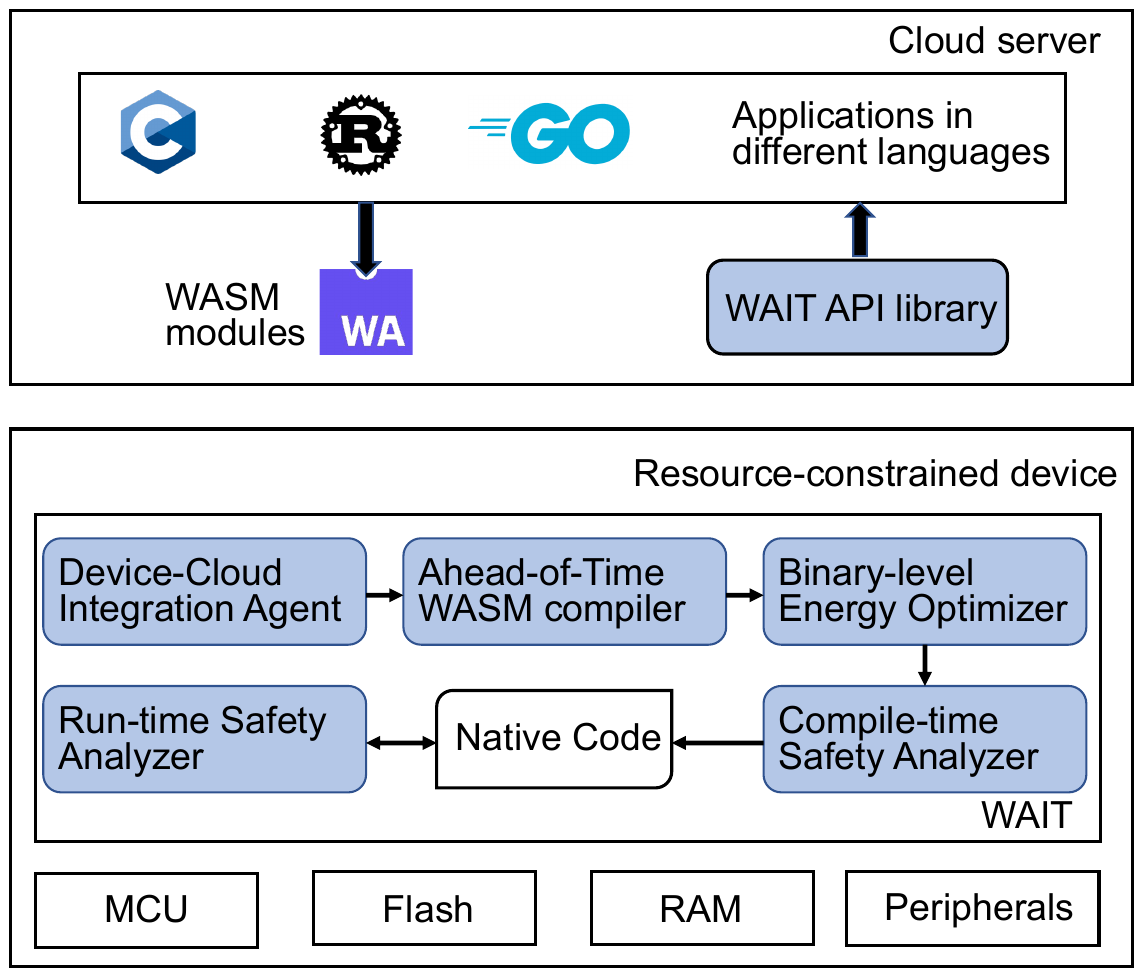}}
	\caption{The workflow of WAIT~\cite{li2022bringing}.}
	\label{figure:WAIT}
\end{figure}

83.3\%(10/12) of the completely designed Wasm runtimes mainly consider the safety feature.
Bosamiya et al. design a safe Wasm runtime without sacrificing efficiency by guaranteeing the sandbox's safety with safe compilers through formal validation~\cite{bosamiya2022provably}. 
Li et al. propose the first Wasm runtime, WAIT, on resource-constrained devices, with safety, efficiency in memory, and energy as the design goal~\cite{li2022bringing}. As shown in Figure~\ref{figure:WAIT}, WAIT offers an API library consisting of IoT APIs for users to develop applications in different languages. Then, the applications are compiled into Wasm modules to be executed in WAIT. WAIT uses AoT compilation to ensure the fast speed of executing Wasm binaries and uses an energy optimizer to save energy. WAIT applies the compile-time and run-time analyzer to check the safety.
Johnson et al. ensure the safety of the proposed Wasm runtime, WaVe, by contributing to the WASI component~\cite{johnson2023wave}. They systematically ensure that WASI, which interacts with the operating system, preserves memory safety and upholds the isolation of access to underlying resources.
Peach et al. utilize the isolation of Wasm runtimes and design a Wasm runtime, eWasm, for microcontrollers~\cite{peach2020ewasm}.
Subramanian et al. design a Wasm runtime, PRIVATION, for general TEE architectures to utilize the safety feature~\cite{subramanyan2023privaton}.

33.3\%(4/12) of the wholly designed Wasm runtimes mainly utilize the lightweight feature.
Gadepalli design a complete runtime, Sledge, utilizing the lightweight Wasm runtime to execute multi-tenant serverless functions on resource-constrained edge systems ~\cite{gadepalli2020sledge}. Sledge uses AoT compiling to convert Wasm binaries into LLVM IR by the Wasm compiler, which is implemented in more than 4,000 lines of Rust code inside Sledge and executes multiple untrusted modules in one process.
Jacobsson et al. utilize the lightweight feature of Wasm runtime on wearable devices~\cite{jacobsson2018virtual}.

Half(6/12) of the entirely designed Wasm runtimes mainly consider the efficiency feature.
Unlike other Wasm runtimes running on OSes, Wasmachine itself is an OS kernel that runs on bare-metal machines and could be faster than native code~\cite{wen2020wasmachine}. Wasmachine runs Wasm modules in the sandboxed kernel thread, invoking system calls as normal functions in Ring 0, which reduces the performance overhead caused by WASI in other Wasm runtimes.
Salim et al. design a Wasm runtime, TruffleWasm, to execute Wasm binaries by the interpreter and JIT optimization, which is the first runtime to execute Wasm binaries in JVM~\cite{salim2020trufflewasm}.
Ménétrey et al. propose an efficient and safe Wasm runtime, WATZ for the TEE~\cite{menetrey2022watz}.
Li et al. propose an IoT OS, ThingSpire OS, based on an originally designed Wasm runtime with a contribution to the Wasm compiler(AoT) component to ensure the execution speed~\cite{li2021thingspire}.

8.3\%(1/12) of the fully completely designed Wasm runtimes primarily concentrate on the compatibility feature.
To be compatible on different Andriod platforms, Wen et al. propose a new Wasm runtime, WasmAndrioid~\cite{wen2022wasmandroid}. Android developers could directly compile source code into Waasm binaries and execute them on various hardware platforms without configuration.

\subsubsection{Adding features for a Wasm runtime}
\begin{figure}[htb]
\centerline{\includegraphics[width=0.90\textwidth]{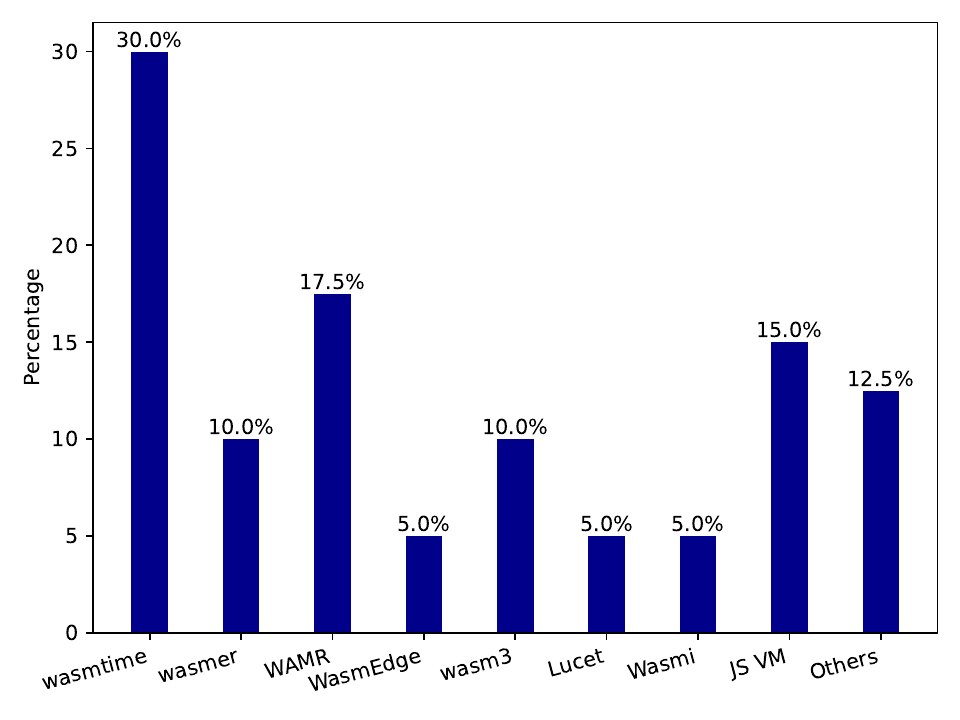}}
	\caption{The fundamental Wasm runtimes used to add features.}
	\label{figure:based_runtime}
\end{figure}

This section is organized by the components modified in the Wasm runtime. 
Figure~\ref{figure:based_runtime} illustrates the foundational Wasm runtimes used for adding features in the 40 articles in \textit{Category A.2}. Due to the consideration that multiple runtimes could be used in one article, these articles are included in the counts for different Wasm runtimes.
As the Wasm runtime supported by the bytecode alliance, \texttt{wasmtime}~\cite{wasmtime} and \texttt{WAMR}~\cite{WAMR} attract the attention of half of the researches in \textit{Catgory A.2}. Moreover, \texttt{wasmtime} is the most widely used runtime. What's more, 82.5\% of the researchers use standalone Wasm runtimes, including \texttt{wasmtime}, \texttt{wasmer}, \texttt{WAMR}, \texttt{WasmEdge}, \texttt{wasm3}, \texttt{Lucet} and \texttt{Wasmi}, the JS VM, such as V8 engine also attracts 15.0\% of the researchers.

\begin{figure}[htbp]
    \centering
    \begin{minipage}[b]{0.40\textwidth}
        \centering
        \includegraphics[width=\textwidth]{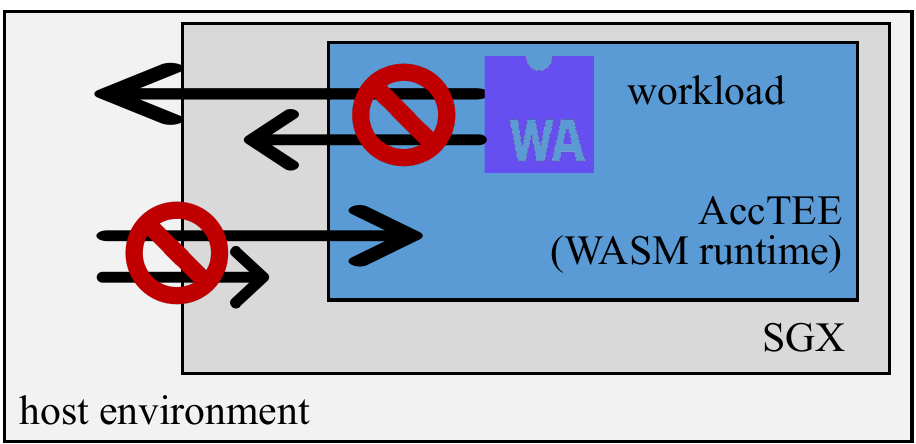}
        \caption{Overview of AccTEE~\cite{goltzsche2019acctee}.}
        \label{figure:AccTEE}
    \end{minipage}
    \hspace{0.02\textwidth}
    \begin{minipage}[b]{0.40\textwidth}
        \centering
        \includegraphics[width=\textwidth]{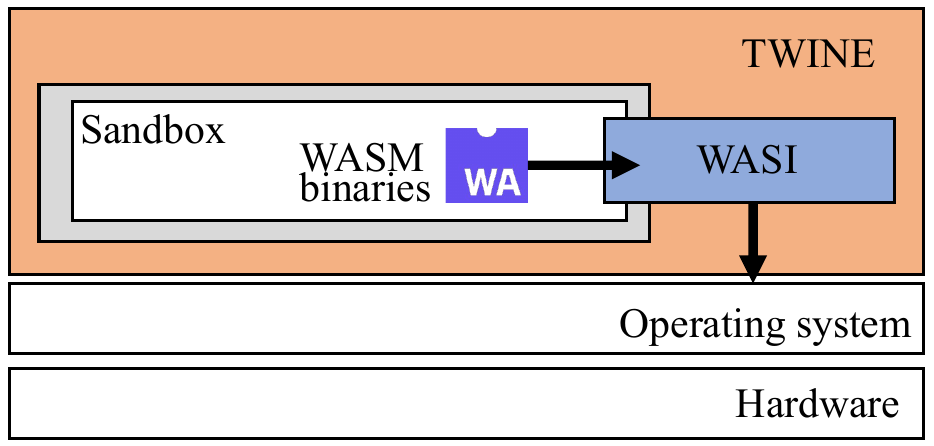}
        \caption{The architecture of TWINE~\cite{menetrey2021twine}.}
        \label{figure:twine}
    \end{minipage}
\end{figure}

57.5\%(23/40) of the articles in \textit{Category A2} contribute to the runtime environment of a Wasm runtime.
Goltzsche et al. design a Wasm runtime, AccTEE, for remote computation, based on the \texttt{Node.js runtime}~\cite{goltzsche2019acctee}. As shown in Figure~\ref{figure:AccTEE}, AccTEE is applied inside the Intel SGX enclave, which is a kind of TEE (trusted execution environment), and utilizes the two-way sandbox of the Wasm runtime to protect the execution process and data due the untrust between infrastructure providers and workload providers. By mainly contributing to the runtime environment component, AccTEE computes the resource usage based on three factors: 1) the number of Wasm instructions, 2) the size of the allocated memory, and 3) the number of I/O operations. 
Liu et al. utilize the sandbox feature of the Wasm runtime, \texttt{WAMR}, which is used as a multi-tenant environment~\cite{liu2021aerogel}.
The optimization of Wasm introduces safety issues related to linear memory. Specifically, it introduces the possibilities by modifying metadata containing memory structure details. Malicious users could utilize heap memory overflow in executing Wasm applications to access arbitrary memory addresses and perform various malicious operations. To ensure the safety of executing Wasm binaries, Song et al. propose that the metaSafer shadow metadata from Wasm linear memory into JS virtual machines and perform metadata verification at a fast speed~\cite{song2023metasafer}.
With the similar purpose of Song et al. to protect Wasm linear memory, Lei et al. propose the first Wasm runtime with memory isolation, PKUWA, based on \texttt{wasmtime}~\cite{wasmtime}. Unlike Song et al., who shadow the metadata in linear memory, PKUWA protects the linear memory at function-level with trivial memory cost and small execution speed cost~\cite{lei2023put}.
Zhang et al. also focus on the safety of linear memory during Wasm execution~\cite{zhang2023vmcanary}. They present a framework, VMCANARY, to protect the memory using the canary approach.
Zhao et al. introduce \texttt{WAMR}~\cite{WAMR} into serverless computing based on the isolated environment the runtime provided and modified the runtime library~\cite{zhao2023reusable}.
M{\'e}n{\'e}trey et al. update the Wasm runtime, \texttt{TWINE}, with the modification of two trusted primitives of attestation, including generate and verify, to extend \texttt{TWINE} into various TEEs~\cite{menetrey2024holistic}.
Geller et al. design a Wasm runtime, Wasm-precheck, based on \texttt{wasmtime}~\cite{wasmtime}. They first extend the Wasm specification to a superset and then use Wasm-precheck to do additional static reasoning instead of the dynamic safety checks, which improve the performance of the Wasm runtime~\cite{geller2024indexed}.
Nie{\ss}en et al. design a Wasm runtime based on \texttt{V8 engine}~\cite{v8} for the Node.js applications with the shared code cache among multi-process to improve the performance~\cite{niessen2020insights}.
Zhu et al. propose a lightweight task scheduling framework for \texttt{WasmEdge}~\cite{WasmEdge} to offload vital Wasm binaries to the edge, with the reduction of memory consume~\cite{zhu2022lawow}.

22.5\%(9/40) of the articles in \textit{Category A2} contribute to the WASI part of a Wasm runtime.
As shown in Figure~\ref{figure:twine}, Ménétrey et al. design TWINE, a lightweight Wasm runtime based on \texttt{WAMR}~\cite{WAMR}, with the modification of the original WASI component to translate the WASI operations to systems calls or TEE libraries~\cite{menetrey2021twine, menetrey2023comprehensive}. In order to protect the data being accessible outside the enclave, TWINE maps the file operating to the Intel-protected file system and utilizes the sandbox nested in the TEE.
Kohn et al. utilize the safety feature provided by the Wasm runtime's sandbox to protect the host system from being influenced by smart contracts from malChadhaicious vendors. Moreover, they extend the API part to interact with the underlying storage system~\cite{almstedt2023contractbox}.
Chadha et al. propose the first Wasm runtime, MPIWasm, for HPC applications based on \texttt{wasmer}~\cite{wasmer, chadha2023exploring}. MPIWasm extends the WASI part to provide the interaction layer between Wasm binaries and the underlying host1 MPI library. 
Sebrechts et al. extend the WASI component in \texttt{wasmtime}~\cite{wasmtime} to enable their runtime being used as Kubernetes controllers~\cite{sebrechts2022adapting}.
Abbadini et al. propose a permission-checking framework with eBPF programs for the WASI part in different Wasm runtimes~\cite{abbadini2023poster}. Wasm runtimes are generally expected to perform security checks to prevent accessing arbitrary locations through host calls in WASI. However, this approach is error-prone and is poor at granularity. Abbadini et al. replace the traditional checking with eBPF programs at fine-grained for each module~\cite{abbadini2023poster}.

17.5\%(7/40) of the articles in \textit{Category A2} contribute to Wasm compiler of a Wasm runtime.
In order to solve the two issues: 1) the code of the IoT applications cannot immigrate seamlessly in heterogeneous platforms, and 2) the code could not be fully portable, Li et al. utilize the compatibility feature of a designed Wasm runtime, WiProg. And they utilize the JIT compiler in \texttt{wasmer} and the interpreter in \texttt{wasm3}~\cite{li2021wiprog} to execute Wasm binaries.
Moron et al. introduce the JIT compiling into \texttt{wasm3}~\cite{wasm3} to alleviate the efficiency limit caused by interpreter~\cite{moron2023support}.
M{\"a}kitalo et al. optimize the implementation of dynamic linking Wasm modules based on \texttt{wasmtime}~\cite{wasmtime} to improve the startup speed~\cite{makitalo2021bringing}.
Gu et al. provide a compiler verifier based on the JIT compiler in \texttt{wasmtime}~\cite{wasmtime, gu2023constant}.
Kolosick et al. modify the Wasm compiler in \texttt{Lucet}~\cite{Lucet} to remove the performance cost of the systems relying on sandboxing mechanism in \texttt{Lucet}~\cite{kolosick2022isolation}.
Narayan et al. propose a compiler framework to protect Wasm from Spectre attacks based on \texttt{Lucet}~\cite{Lucet, narayan2021swivel}.

17.5\%(7/40) of the articles in \textit{Category A2} focus on the Wasm interpreter.
In contrast to other runtimes like TruffleWasm~\cite{salim2020trufflewasm}, Wasmachine~\cite{wen2020wasmachine}, and WAIT~\cite{li2022bringing}, Titzer et al. employ a space-efficient and rapid in-place Wasm interpreter to prioritize efficiency. This choice is made considering that for JIT compilers, both compilation time and memory usage profoundly impact application startup time~\cite{titzer2022fast}.
Pop et al. extend the Wasm interpreter of WASMI by adding serialization, deserialization pausing, etc., for the secure enclave service migration~\cite{pop2022towards}.
Watt et al. propose a monadic fast Wasm interpreter based on \texttt{wasmtime}~\cite{wasmtime} to be used as a fuzzing oracle~\cite{watt2023wasmref}.
Marques et al. extend the Wasm interpreter~\cite{wasminterpreter} with symbolic facilities to analyze the execution of Wasm binaries~\cite{marques2022concolic}.
Nurul-Hoque et al. develop an interpreter architecture for \texttt{wasm3}~\cite{wasm3} to separate the runtime state from the host environment, enabling \texttt{wasm3} supporting live-migration~\cite{nurul2021nomad}.

\begin{tcolorbox}
\textbf{Summary of answers to RQ2-1:} \\
\textbf{(1)} More than half of the papers related to Wasm runtimes design a Wasm runtime entirely or based on existing Wasm runtimes, which shows that the design and utilization of Wasm runtimes currently attract the attention of the majority of researchers. Instead of designing an entire Wasm runtime, most researchers design a new runtime based on the existing Wasm runtimes, with wasmtime being the most used one.\\
\textbf{(2)} The majority of researchers primarily consider the runtime environment component when designing Wasm runtime, with the Wasm interpreter being the least considered component. This is mainly because the runtime environment component encompasses all functionalities besides Wasm instruction execution and the Wasm System Interface, such as memory allocation and resource usage calculation. Furthermore, only a minority of Wasm runtimes support execution through interpreters. Hence, the Wasm interpreter is the least considered component.\\
\textbf{(3)}The safety feature is researchers' most commonly utilized characteristic of Wasm runtimes, followed by the efficient feature. The potential of Wasm applications can be explored in various scenarios by designing Wasm runtime and ensuring its safety and efficiency.
\end{tcolorbox}
\subsection{RQ2-2:Wasm runtime testing}

Instead of directly contributing to the design of the Wasm runtime itself, 33.7\% (33/98) of researchers contribute to testing Wasm runtimes, promoting the development of the runtime by identifying shortcomings.

\subsubsection{Performance testing}
\begin{figure}[htb]
\centerline{\includegraphics[width=0.85\textwidth]{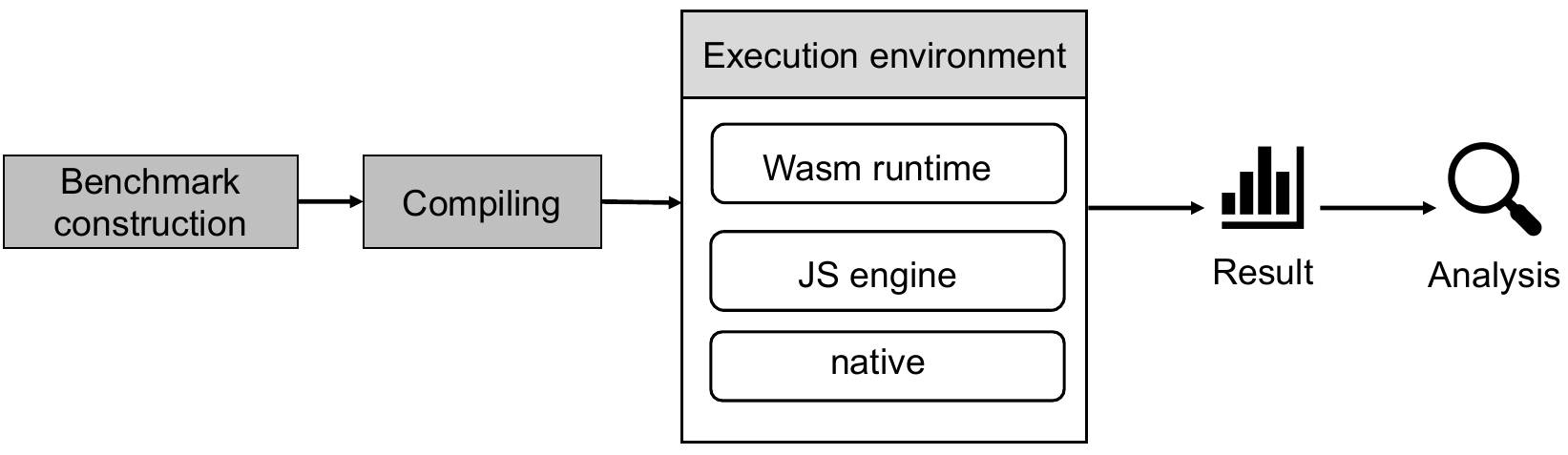}}
	\caption{The general workflow of performance and energy consumption testing.}
	\label{figure:testing_workflow}
\end{figure}

As shown in Figure~\ref{figure:testing_workflow}, we summarize the performance and energy consumption testing workflow of Wasm runtimes. First, the researchers construct the benchmark in different languages and then compile the benchmark into Wasm binaries or native code. Second, the compiled benchmarks are deployed in different execution environments, including Wasm runtimes, JS engines, and native environments. For example, directly compiling high-level language benchmarks into native code and deploying them directly in a Linux environment. Finally, the researchers collect and analyze the execution result.

\begin{center}
\begin{longtable}{ p{0.25\textwidth} | p{0.25\textwidth} | p{0.40\textwidth}}
\caption{The benchmarks used when testing Wasm runtimes.}
\label{benchmark_test_Wasm_runtime} \\
\toprule
\textbf{Benchmark} & \textbf{Usage} & \textbf{Description}\\
\midrule
WABench~\cite{wang2022far} & performance testing & A benchmark contains 50 programs from JetStream2, MiBench, PolyBench, and the manually built part.\\
\cline{1-3}
Ostrich Benchmark Suite~\cite{oliveira2020analysis, van2022comparing} & performance testing, erergy consumption testing & A test suite comprises 12 algorithms with the same implementation in JS and C.\\
\cline{1-3}
Wasm-MB and Wasm-BM~\cite{aliyev2023analysis} & performance testing & The WebAssembly Micro benchmarks (WASM-MB) and WebAssembly Benchmarks (WASM-BM) are often used in Wasm performance testing and cover various use cases. \\ 
\cline{1-3}
SPEC benchmarks~\cite{hockley2022benchmarking, jangda2019not} & performance testing & A test suite of micro-benchmarks and a macro-benchmark, both written in Rust. \\
\cline{1-3}
Fibonacci benchmark~\cite{mohan2022comparative} & performance testing & A program to calculate the Fibonacci of a number.\\ 
\cline{1-3}
Prime benchmark~\cite{mohan2022comparative} & performance testing & A program to calculate whether a given number is prime or not.\\
\cline{1-3}
Stotoglou's benchmark~\cite{stotoglou2023comparative} & performance testing & A test suite includs numerical computation, classification and image processing.\\
\cline{1-3}
Benchmark.js~\cite{kyriakou2022complementing} & performance testing & A benchmark library with high-resolution timer support, providing statistically significant results.\\
\cline{1-3}
XR-related framework or library~\cite{liu2023demystifying} & performance testing & A test suite consists of tasks of Web XR.\\
\cline{1-3}
PolyBench C benchmark~\cite{wang2021empowering, spies2021evaluation, wang2021characterization, szewczyk2022leaps} & performance testing & A test suite compiled from 30 widely-used C benchmarks.\\
\cline{1-3}
Mendki's benchmark~\cite{mendki2020evaluating} & performance testing & A test suite includes different applications, such as compute-intensive tasks, memory-intensive tasks, file I/O intensive, and a simple image classification - machine learning application. \\
\cline{1-3}
Part of the Computer Language Benchmarks Game (CLBG)~\cite{wagner2023energy, de2021runtime} & performance testing, energy consumption testing & A test suite implements ten computational problems in up to 26 languages. The authors use 3 of the 10. \\
\cline{1-3}
Macedo's benchmark~\cite{de2021runtime} & performance testing, energy consumption testing & A test suite mainly consists of sorting computation.\\
\cline{1-3}
LLVM test suite~\cite{jiang2023revealing}. & performance testing & A test suite contains various programs to evaluate the performance of LLVM compilation.\\
\cline{1-3}
The SuiteSparse Matrix Collection~\cite{sandhu2018sparse} & performance testing & A set of sparse matrix benchmarks from real life.\\
\cline{1-3}
Yan's benchmarks~\cite{yan2021understanding} & performance testing & The test suite consists of three parts: 1)41 widely used C benchmarks, 2)9 hand-written JS programs, and 3)3 real-world programs in Wasm and JS.\\
\cline{1-3}
Scheidl's benchmarks~\cite{scheidl2020valent} & performance testing & A test suite consists of computing algorithms from two sources.\\
\cline{1-3}
Zheng's benchmarks~\cite{zheng2020vm} & performance testing & A test suite contains smart contracts from the real world or manually written.\\
\cline{1-3}
Kjorveziroski's benchmarks~\cite{kjorveziroski2023webassembly} & performance testing & A test suite from micro-benchmarks and real world programs.\\
\cline{1-3}
Pham's benchmarls~\cite{pham2023webassembly} & performance testing & A benchmark consists of compute-intensive workloads centered around computing Message Digest Method 5 (MD5) hashes.\\
\cline{1-3}
Junior's benchmarks~\cite{junior2020webassembly} & performance testing & A test suite contains digital image processing applications.\\
\cline{1-3}
Macedo's benchmarks~\cite{de2022webassembly} & performance testing, energy consumption testing & Mirco benchmarks form C, and real-world benchmarks are built from WasmBoy and PSPDFKit.\\
\cline{1-3}

Rosetta Code platform~\cite{pockstaller2023comparing} & energy consumption testing & A website dedicated to programming exercises providing implementations of common algorithms in various programming languages.\\
\cline{1-3}

WasmFuzzer's test suite~\cite{jiang2022wasmfuzzer} & bug detection & Wasm programs generated automatically by WasmFuzzer.\\
\cline{1-3}
WADIFF's test suite~\cite{zhou2023wadiff} & bug detection & Wasm binaries generated based on the Wasm specification.\\ 
\cline{1-3}
WRTester's test suite~\cite{cao2023wrtester} & bug detection & A test suite consists of Wasm programs by disassembling and assembling real-world Wasm binaries.\\ 
\bottomrule
\end{longtable}
\end{center}

\textbf{Benchmark.}
As shown in Table~\ref{benchmark_test_Wasm_runtime}, we summarize the benchmarks used for testing Wasm runtimes.

\textit{Some articles apply existing benchmarks to test the performance of Wasm runtimes by compiling the high-level programs in existing benchmarks into Wasm binaries.}
Ostrich Benchmark Suite consists of 12 algorithms~\cite{oliveira2020analysis}, such as computing the optimal alignment of two protein sequences, computing n queen problem, calculating particle potential and relocation within a 3D space, etc. Both the C and JS versions in the same implementation are provided.
SPEC benchmarks were first proposed by Jangda et al. in 2019~\cite{jangda2019not} for the Web. Hockley et al. extended SPEC benchmarks to the non-Web environment in 2022~\cite {hockley2022benchmarking}.
Stotoglou et al. build the benchmark based on similar works considering the availability and compatibility with three types of tasks~\cite{stotoglou2023comparative}. 
Kyriakou et al. utilize Benchmark.js, a benchmarking library~\cite{kyriakou2022complementing}.
In order to meet the need of testing Web XR performance, Liu et al. introduce three XR-related tasks, including loading and rendering 3D content, sampling images, and tracking objects~\cite{liu2023demystifying}.
Wang et al. compile 30 C programs in the PolyBench C into Wasm binaries~\cite{wang2021empowering}. 
Moreover, M{\"a}kitalo et al. and Wang et al.~\cite{wang2021characterization} also apply the PolyBench to compare the performance among different Wasm runtimes~\cite{makitalo2021bringing}.
Wagner et al. introduce part of the computer language benchmarks as it has programs written in different languages with the same implementation~\cite{wagner2023energy}.
Macedo et al. choose sorting programs from the programming chrestomathy repository, Rosetta Code~\cite{de2021runtime}. Two intensive benchmarks compatible with WebAssembly are incorporated, namely fannkuch-redux and fasta, sourced from the Computer Language Benchmarks Game (CLBG).
Jiang et al. introduce the LLVM test suite~\cite{jiang2023revealing}. They select 141 C/C++ programs from the various benchmarks for testing LLVM compilation.
Sandhu et al. collect close to 2,000 test cases from the SuiteSparse Matrix Collection and classify the benchmarks into seven categories according to the number of non-zeros~\cite{sandhu2018sparse}.
Besides, Pham's and Junior's benchmarks~\cite{pham2023webassembly, junior2020webassembly} are also extracted from existing benchmarks.

\textit{The rest part of the articles use the benchmarks built by the authors.}
The Fibonacci benchmark contains the program of calculating Fibonacci sequences of a number~\cite{mohan2022comparative}. The sequence is generated by adding the previous two numbers, which is a good choice for testing recursive calls. The prime benchmark is used in the same article by Tushar et al. The program needs to run a loop to judge whether the given number is prime, which is a good choice for testing iterative programs.
Mendki develops a test suite that includes different applications written in Rust~\cite{mendki2020evaluating}. Inside, the compute-intensive tasks are derived from the wasmer test suite.
Yan et al. build the benchmarks in three ways: hand-writing, real-world collection, and extracting from existing benchmarks. They write 9 JS programs and collect three Wasm binaries and JS applications from the real world~\cite{yan2021understanding}.
Scheidl et al. build the benchmarks from two source~\cite{scheidl2020valent}. Some benchmarks are randomly selected from PolyBench C, including nussinov, floydwarshall, jacobi1d, and correlation. The rest of the algorithm programs are written by themselves, including Fibonacci sequence computation, Mandelbrot set calculating, and the Pollard Rho integer factorization algorithm.
Zheng et al. build 13 smart contracts manually, containing four kinds of operations: simple operations, arithmetic operations, block status-related operations, and hashing operations~\cite{zheng2020vm}. Moreover, they also collect 13 Solidity smart contracts from the real world.
Besides, Kjorveziroski et al. also select benchmarks from the real world~\cite{kjorveziroski2023webassembly}.
Macedo et al. build the benchmarks from C benchmarks and the real world, including the WasmBoy benchmark and PSPDFKit Benchmark~\cite{de2022webassembly}.

\begin{figure}[htb]
\centerline{\includegraphics[width=0.75\textwidth]{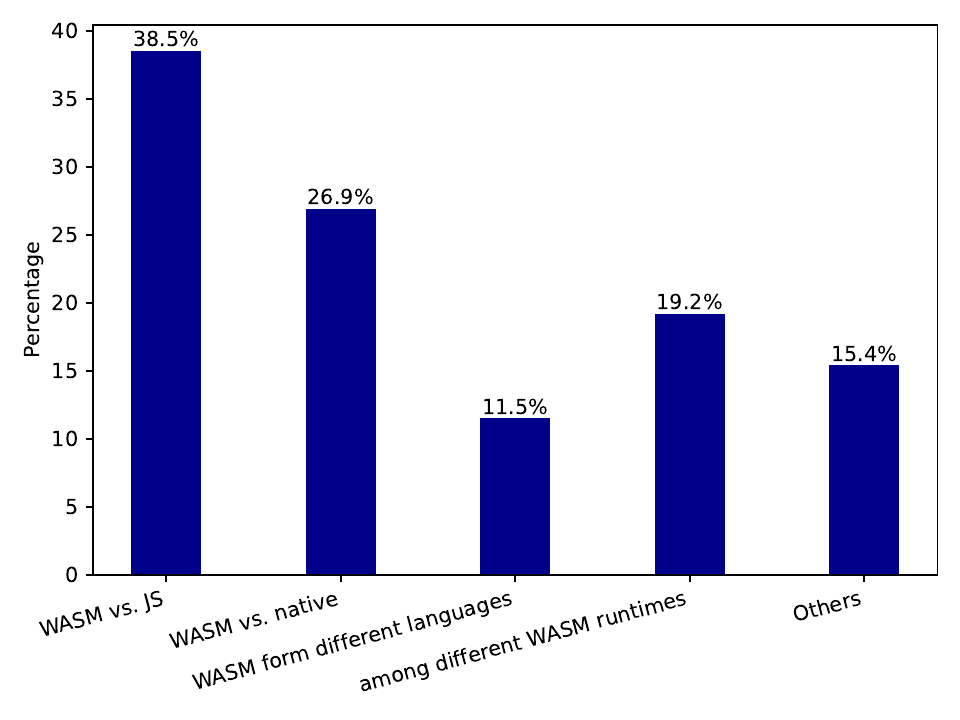}}
	\caption{The method percentage in performance testing of Wasm runtimes.}
\label{figure:method_performance_Wasm_runtime}
\end{figure}

\textbf{Method.}
In this section, we compare the methods used in performance testing of Wasm runtimes. As shown in Figure~\ref{figure:method_performance_Wasm_runtime}, the methods the articles used are collected. The most widely used method is to compare the performance of Wasm, JS, and native code. As Sandhu et al.'s~\cite{sandhu2018sparse}, Mohan et al.'s~\cite{mohan2022comparative}, Spies et al.'s~\cite{spies2021evaluation} and Wang et al's~\cite{wang2022far} work cover two methods, they are calculated in both methods. 

\textit{38.5\% (10/26) of the articles in performance testing of Wasm runtimes focusing on comparing Wasm and JS.} As Wasm was first proposed to improve the performance of JS, whether the execution of Wasm is indeed superior to JS remains the main focus of attention.
Oliveira et al. explore Wasm's capability to enhance JS's performance~\cite{oliveira2020analysis}. They use the Ostrich Benchmark Suite, which provides programs with the same functionality in both C and JS. Thus, the Wasm counterpart could be generated from C. Finally, they compare the execution time of JS and Wasm in 8 of 12 algorithms in the benchmark, as the rest cannot be compiled into Wasm binaries due to compatibility. All the execution is on Raspberry Pi. In all cases, Wasm is faster than JS, especially in reducing battery-power consumption. And same result is found by Macedo et al~\cite{de2022webassembly}.
Macedo et al. also find that Wasm performs better in the performance and energy consumption in most cases, and the larger the program input, the more pronounced the advantage of Wasm~\cite{de2021runtime}.
As sparse matrix-vector multiplication is a representative computation of compute-intensive programs, Sandhu et al. use sparse matrices to test the performance ~\cite{sandhu2018sparse}.
Yan et al. perform the testing based on comprehensive types of applications, including real-world programs, hand-written programs, and programs from other existing benchmarks~\cite {yan2021understanding}. 
Kyriakou et al. conducts stress tests to compare the performance of JS and Wasm~\cite{kyriakou2022complementing}. In Web browsers, Wasm in a single thread performs 2-4x faster than the pure JS implementation. 
Tushar et al. compare the performance of Wasm and JS on Chrome and Firefox~\cite{mohan2022comparative}. The Wasm binaries are compiled from the C, Rust, and Go versions of calculating the Fibonacci sequence of a given starting number. Wasm performs better than JS when the iterating of Fibonacci becomes more extensive, and Wasm in Firefox is faster than in Chrome. Compared to the Wasm binaries compiled from Rust or Go, Wasm binaries from C/C++ perform better.
Through comparing the execution of JS and Wasm browsers, Wang studied how the browsers optimize the execution process of Wasm binaries~\cite{wang2021empowering}. However, although the JIT optimization could accelerate JS significantly, it fails for Wasm. Moreover, Wasm consumes much more memory than JS, which is a good opportunity for the developers to improve. 
Stotoglou et al. test the performance of Wasm and JS in different browsers to compare the performance of different types of tasks across different browsers~\cite{stotoglou2023comparative}. Wasm is found to be faster in all types of tasks. Performance largely depends on the input data, algorithms, hardware, and the browser's version. However, different browsers have their own optimization strategies; the V8 engine usually performs the best.

\textit{Moreover, 26.9\% (7/26) of the articles compare Wasm and native code, marking it as the second significant focus.}
Mendki et al. find that Wasm runtime is not as fast as expected, hindering the application of Wasm in serverless computing~\cite{mendki2020evaluating}.
And Szewczyk et al. investigate the performance impact arising from the bounds-checking mechanisms in WebAssembly~\cite{szewczyk2022leaps}. 
Jangda et al. first propose the SPEC benchmarks by extending BROWSIX~\cite{jangda2019not}, ensuring the running of the Unix programs in Wasm without modifications. Second, they offer an automatically detailed timing collection and hardware performance counting tool, BROWSIX\_Wasm, to conduct fine-grained performance testing. Third, they conduct the first comprehensive performance analysis of Wasm in 2019, finding a significant gap between Wasm and native code. The reasons are: 1) Wasm instructions entail a higher frequency of load and store operations; 2) Wasm instructions encompass more branches owing to safety checks; and 3) Wasm results in increased L1 instruction misses due to the higher volume of instructions.
Junior et al. verify the promising future of Wasm as it performs similar performance of native code~\cite{junior2020webassembly}.

\textit{11.5\% (3/26) of the articles focus on the Wasm binaries' performance compiled from different high-level languages.}
Aliyev et al. use the Web Assembly Micro benchmarks (Wasm-MB) and Web Assembly Benchmarks (Wasm-BM) benchmark suites to test the runtimes's performance ~\cite{aliyev2023analysis}. Their testing environment is oriented towards cloud providers, including Google Cloud Platform (GCP), Amazon Web Services (AWS), and Microsoft Azure. They believe that the execution time between 15ms and 20ms is a good performance, while the execution time between 20ms and 30ms is poor. Moreover, the Wasm runtime performs well in their view.
Wagner et al. utilize three of the ten algorithms from the Computer Language Benchmarks to assess the impact of various high-level languages, including C, Rust, Go, and JS, on Wasm execution performance~\cite{wagner2023energy}. They find that C and Rust could be better choices to save energy in IoT devices, and different Wasm runtimes exhibit preferences for different high-level languages.

\textit{19.2\% (5/26) articles dive into the performance in different Wasm runtimes.}
Hockley et al. compare the performance with different Wasm compilers in \texttt{wasmer}~\cite{hockley2022benchmarking, wasmer}. They use \texttt{wasmer} to execute the SPEC benchmarks (micro and macro benchmarks written in Rust). After comparing the performance of three Wasm compilers (singlepass, cranelift, LLVM) used in \texttt{wasmer}, they find that cranelift and LLVM perform better, since basically, singlepass does not perform any optimization. Although LLVM performs better than cranelift in some test cases, LLVM costs a longer-time JIT process than cranelift with a more complex compilation.
Jiang et al. first propose a testing approach, WarpDiff, based on the differential testing framework~\cite{jiang2023revealing}. That is to say, for each test case, the execution ratio of different Wasm runtimes obeying a fixed value is considered to be normal, and the execution time that does not obey a fixed ratio is considered to be anomalous, which needs to be further analyzed. Second, they compare the performance of five standalone Wasm runtimes, including \texttt{wasmer}~\cite{wasmer}, \texttt{wasmtime}~\cite{wasmtime}, \texttt{WAMR}~\cite{WAMR}, \texttt{WasmEdge}~\cite{WasmEdge} and \texttt{wasm3}~\cite{wasm3} on more than 100 test cases extracted from the LLVM test suite. Third, seven performance issues are detected, all confirmed by the runtimes' developers.
Scheidl et al. proposed an optimization through valent-blocks and getting performance improvement over other Wasm runtimes~\cite{scheidl2020valent}.
Kjorveziroski et al. compare the performance of \texttt{wasmtime}~\cite{wasmtime}, \texttt{wasmer}~\cite{wasmer} and \texttt{WasmEdge}~\cite{WasmEdge} to judge which is the best in serverless computing~\cite{kjorveziroski2023webassembly}.

\textit{The rest 4 articles use other methods.} 
Zheng et al. compare the performance of smart contracts in Solidity and the counterpart in Wasm binaries, finding that the overhead introduced by the gas metering and switch of EEI (Ethereum environment interface) methods constrain the performance of Wasm smart contracts in eWasm engines~\cite{zheng2020vm}.
Pham et al. commend Wasm runtimes as a resolution of IoT applications, especially \texttt{Wasmedge}~\cite{pham2023webassembly, WasmEdge}.
Sunarto et al. do not directly run Wasm binaries to compare the runtimes' performance. Instead, they compare the performance, memory usage, and energy usage between Wasm and JS of the website development~\cite{sunarto2023systematic}. They use the PRISMA methodology to systematically review the papers comparing Wasm and JS performance. After analyzing no more than 30 works, they find that, in energy consumption, Wasm performs better; however, in memory usage, JS performs better. As for performance, Wasm and JS perform differently in various tasks. Wasm shows faster speed in lightweight applications and numerical calculations, while JS shows rapid speed in heavy applications. Several reasons could influence these results, including devices, browsers, etc.
Liu et al. explore how Wasm could improve Web XR~\cite{liu2023demystifying}. Targeting the XR-related framework or library, they did the first systematic study comparing the performance improved by Wasm in Web XR on five browsers. Although a performance gap exists between Web XR improved by Wasm and the standalone XR, Wasm does accelerate XR in Web.

\textbf{Result analysis.}
By summarizing the results of various papers, we find that Wasm outperforms JS regarding performance and energy consumption~\cite{oliveira2020analysis, de2022webassembly, de2021runtime, kyriakou2022complementing, mohan2022comparative, stotoglou2023comparative}. However, the performance of Wasm compiled from different high-level languages varies~\cite{mohan2022comparative, wagner2023energy}. For instance, C/C++ performs better than Rust and Go. Additionally, Wasm tends to consume more memory than JS~\cite{wang2021empowering}.
However, the performance comparison between Wasm and native is not quite satisfactory~\cite{mendki2020evaluating, jangda2019not}. This is mainly due to the increased number of load and store instructions and safety checks in Wasm~\cite{jangda2019not}. Nonetheless, the performance gap between Wasm and native can still be narrowed~\cite{junior2020webassembly}.
Different Wasm runtimes in browsers or standalone show different performances, such as different optimization strategies~\cite{stotoglou2023comparative}, and exception time consumption ~\cite{jiang2023revealing}.

\subsubsection{Energy consumption testing}

For devices with limited battery capacity, such as smartphones, executing complex web applications can be challenging, making Wasm a potentially better choice. There are 6.1\% (2/33) of the articles mainly test the energy consumption of executing Wasm in the runtimes. 

Hasselt et al. first test the energy consumption in 2022~\cite{van2022comparing}. As shown in Table~\ref{benchmark_test_Wasm_runtime}, they also use the Ostrich Benchmark Suite, the same as Oliveira et al. ~\cite{oliveira2020analysis}. They compare the energy consumption between Wasm and JS on mobile devices with Firefox and Chrome. Wasm saves more energy than JS, and Firefox is better than Chrome.
Pockstaller et al. also compare JS and Wasm in energy consumption in 2023~\cite{pockstaller2023comparing}. They select 19 algorithms from the Rosetta Code platform, each providing implementations in both C and JS. The results are similar to those of Hasselt et al., where JavaScript (JS) also performs better, with specific energy savings ranging from 20\% to 30\%.

\subsubsection{Bug detection}

The Wasm runtime significantly influences the correctness and security of Wasm binaries' execution. About 15.2\% (5/33) of the articles on Wasm runtime testing focus on bug detection in Wasm runtime.

\begin{figure}[htb]
\centerline{\includegraphics[width=0.85\textwidth]{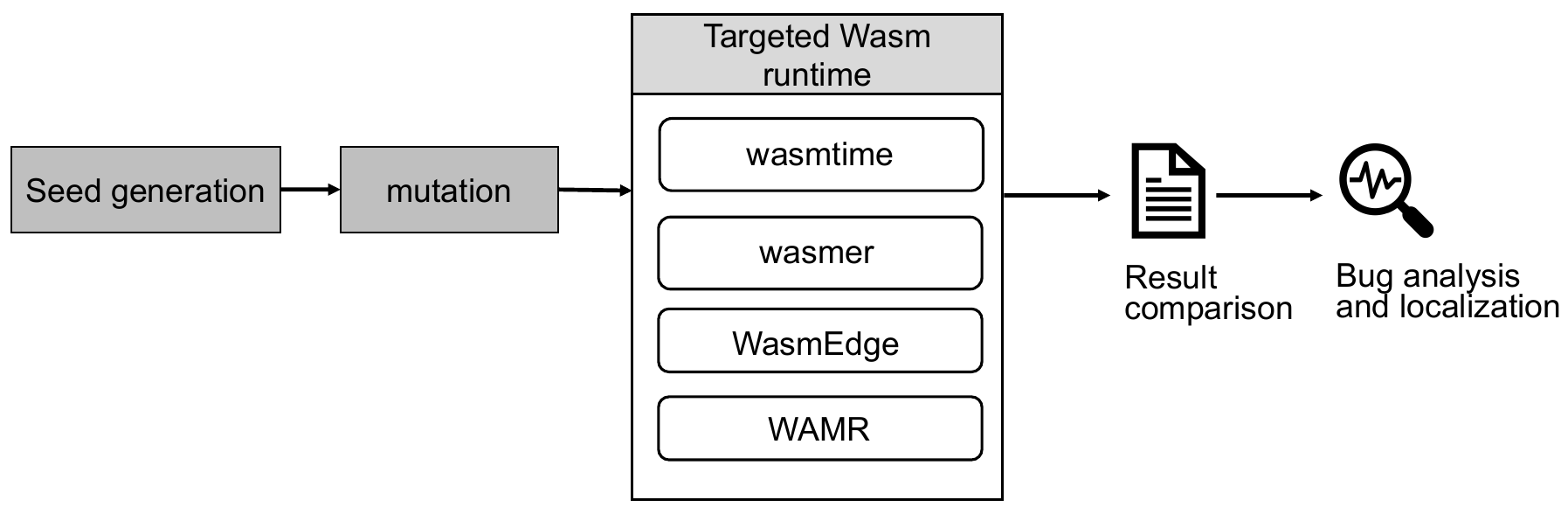}}
	\caption{The general workflow of bug detection.}
\label{figure:bug_detection_workflow}
\end{figure}

As shown in Figure~\ref{figure:bug_detection_workflow}, the researchers to detect bugs first generate seeds based on the Wasm specification or randomly and then mutate the seeds. Second, they deploy the generated Wasm binaries into different Wasm runtimes to execute. Finally, they compare the results from different Wasm runtimes and analyze or locate the bug.

Jiang et al. develop a fuzzing tool, WasmFuzzer, for Wasm runtimes~\cite{jiang2022wasmfuzzer} with automitivally generated seeds and mutation operator at the binary level, performing better than AFL (American fuzzy lop)~\cite{AFL}. As different Wasm runtimes show different bug patterns and have various implementations, the mutation strategy in WasmFuzzer is adaptive to update the priorities of the mutation operators dynamically. Specifically, the mutation operators could be divided into two categories: 1) operators on Wasm instructions, including the instructions' insert, delete, and swap, and 2)other operators, including the add and delete of start, memory sections, etc. Finally, they apply WasmFuzzer and AFL in three Wasm runtimes: \texttt{WAMR}~\cite{WAMR}, \texttt{WAVM}~\cite{WAVM} and \texttt{EOS-VM}~\cite{eos-vm}, which are easier for code instrumentation and computing code coverage. WasmFuzzer performs better than AFL both on unique crashes and code coverage.
Ha{\ss}ler et al. modify the Wasm runtime, \texttt{WAVM}~\cite{WAVM}, to calculate the coverage data for AFL++ Fuzzer and implement lightweight snapshots for the runtime to help detect bugs ~\cite{hassler2021wafl}.

Unlike Jiang et al.'s code coverage guided method, Zhou et al. move their eyes on the Wasm specification~\cite{zhou2023wadiff}. They introduce the Wasm runtimes' differential testing framework, WADIFF, to test seven Wasm runtimes and analyze the bugs' root causes. First, they convert the Wasm specification in natural language into structured DSL format. Second, they design a symbolic execution to generate Wasm binaries. Finally, they apply WADIFF into seven Wasm runtimes, including \texttt{WAMR}~\cite{WAMR}, \texttt{wasmer}~\cite{wasmer}, \texttt{WasmEdge}~\cite{WasmEdge}, etc. Twenty-one bugs are detected, with seven confirmed, resulting from various issues such as different feature support, undefined code, illegal alignment, etc.

Cao et al. devise a runtime-independent root cause identification algorithm that accurately pinpoints bugs using a static instrumentation approach to detect issues at both function and instruction levels~\cite{cao2023wrtester}. Unlike WADIFF generating Wasm binaries based on Wasm specification, the differential testing framework WRTester Cao et al. used produces syntactically correct, semantically rich Wasm programs by disassembling and assembling real-world Wasm binaries. Finally, 33 unique bugs are detected with 25 confirmed, due to memory check, type conversion, etc.

Unlike the methods used in the above three articles, which can be applied in various runtimes and are regarded as general methods, VanHattum et al. narrow the focus to the lowering process of instructions within a Wasm compiler, cranelift~\cite{vanhattum2024lightweight}. They present a framework Crocus by adding concise semantics alongside definitions in ISLE to verify the lowering rules and find two unknown bugs and three found bugs, one of which is a CVE.

\begin{tcolorbox}
\textbf{Summary of answers to RQ2-2:} \\
\textbf{(1)} Currently, testing of Wasm runtimes primarily focuses on performance, bugs, and energy consumption, with performance testing occupying the majority. \\
\textbf{(2)} Researchers choose to build the benchmark themselves or compile the existing benchmarks into Wasm binaries for testing Wasm runtimes. These benchmarks include several kinds of algorithms, such as Fibonacci or some calculations from the real world.\\
\textbf{(3)} Currently, the methods for testing the performance of Wasm runtime are relatively limited. They primarily involve comparisons between different Wasm runtimes, the execution of Wasm binaries compared to JS, and the execution of Wasm binaries compared to native code, among other methods. Despite outperforming JS in performance, Wasm's performance compared to native code often falls short than expected. Moreover, bug detection for Wasm runtime primarily relies on fuzzing and the principles of differential testing.
\end{tcolorbox}
\subsection{RQ2-3:Wasm runtime analysis}
There are 13.3\% (13/98) that analyze the security issues and the future directions or conduct an empirical study of Wasm runtimes.

\subsubsection{Empirical study}
5.1\% (5/98) of the articles apply empirical studies on the Wasm runtimes.

Two articles focus on the positive aspects of Wasm runtimes, including the promising features or usage cases. Titzer provide an empirical study of the single-pass compilers in Wasm runtimes~\cite{titzer2023whose}. Kjorveziroski et al. focus on the practicality of using Wasm runtimes in serverless by describing the current developments~\cite{kjorveziroski2022evaluating}.

The other three articles focus on the issues that hinder the development of Wasm runtimes. Zhang et al.~\cite{zhang2023characterizing} and Wang et al.~\cite{wang2023comprehensive} provide empirical studies for bugs in Wasm runtimes, both constructing taxonomy for the bugs. Besides, Zhang et al. dive deep and summarize the fix strategies and propose a pattern-based bug detection framework based on the statistics of reported bugs and find 60 unknown bugs, with 13 confirmed and 9 fixed.
By analyzing the reported issues from GitHub and Stack Overflow, Waseem et al. find that 28.16\% of the issues that Wasm application developers encountered are introduced by Wasm runtimes~\cite{waseem2023understanding}.

\subsubsection{Discussion about future directions}
6.1\% (6/98) of the articles provide insights and discussions on the future development directions of Wasm runtimes. These papers are forward-looking studies in the application domains of Wasm runtimes, presenting ideas for various development directions.

Wasm runtimes provide a portable execution environment for the execution of Wasm binaries, which attracts the researchers' attention.
M{\"a}kitalo et al. proposed a vision using WebAssembly (Wasm) to implement lightweight containers for the IoT domain in 2021~\cite{makitalo2021webassembly}. This leverages the poerability features of Wasm, allowing the code of IoT applications to be easily deployed and migrated across any IoT device in the system.
Hoque et al. also utilize the compatibility and portability of Wasm and propose four proposals for using Wasm runtimes in Edge offloading computimg~\cite{hoque2022webassembly}.
M{\'e}n{\'e}trey et al. proposed using Wasm to solve the incompatibility between large data centers and user devices~\cite{menetrey2022webassembly}.

Besides, others focus on Wasm's high efficiency. Wallentowitz utilizes the high efficiency and isolation of the runtimes provided for embedded systems~\cite{wallentowitz2022potential}. They demonstrate Wasm's viability and discuss its challenges, such as application management, minimum memory requests conflicting with embedded system resource constraints, etc. 
Joshi analyzes how to use Wasm outside the Web, including in docker, cloud, microservers, etc~\cite{joshianalysis}. 
Kakati et al. regard the Wasm as an ideal resolution for applications in edge computing due to its compact format and the high-efficiency execution environment~\cite{kakati2023webassembly}.

\subsubsection{Security analysis}
There are 2.0\% (2/98) articles focus on the security issues in Wasm runtimes.
Yang et al. analyze the security issues in the Wasm runtimes that execute smart contracts in Wasm by the tool Seraph~\cite{yang2020seraph}. Seraph develops an abstraction of the interaction layer to the underlying blockchain platform and describes the security issues by the symbolic semantic graph. 
Puddu et al. execute the confidential code as compiling to Wasm binaries in \texttt{WAMR}~\cite{WAMR} in the TEE environment, finding that TEE leaks most IR instructions (Wasm binaries here) with the revealing of the confidential code, questing the commercial resolutions of TEEs+Wasm~\cite{puddu2022lack}.

\begin{tcolorbox}
\textbf{Summary of answers to RQ2-3:} \\
\textbf{(1)} Researchers primarily focus on both the promising aspects and the obstacles hindering the development of Wasm runtimes. This includes assessing the practicality of Wasm offered by these runtimes, identifying the challenges introduced by Wasm runtimes that Wasm application developers encounter, and categorizing the types of bugs present in Wasm runtimes. \\
\textbf{(2)} The researchers also propose promising future applications, such as using Wasm runtimes as lightweight containers in IoT and applying them to address compatibility issues between large data centers and user devices.\\
\textbf{(3)} Ensuring the security of Wasm runtime is crucial. Despite its widespread adoption, researchers have questioned existing solutions for Wasm runtime security by analyzing security issues in smart contracts and TEE environments.
\end{tcolorbox}

\begin{tcolorbox}
\textbf{Summary of answers to RQ2:} \\
The existing articles on Wasm runtime research mainly focus on three aspects: the design of Wasm runtime, testing of Wasm runtime, and analysis of Wasm runtime, with the design of Wasm runtimes for various scenarios taking the main proportion.
\end{tcolorbox}

\section{Future research directions}\label{sec:future}
This section discusses problems that exist in current articles and the future directions.

\subsection{Wasm runtime design}
53.1\% of the articles explore the design of Wasm runtimes. Currently, the design of Wasm runtime has covered various fields and has gradually matured, but there are still areas for improvement.

As for the \textbf{Wasm runtime design for a specific scenario}, we provide the following advice:

1) As to applying Wasm into an embedded system, the minimal memory request of 64KB conflicts with the restrict resource limitation formulated in embedded systems~\cite{wallentowitz2022potential}. The Wasm runtime designers and the specification formulator could provide the configurable requests of the linear memory to adapt to this scenario.

2) As to applying Wasm runtimes to executing smart contracts, especially in Ethereum, although the execution of instructions is high-performance, the interaction between Wasm binaries and the blockchain through the EEI methods hinders the high efficiency~\cite{zheng2020vm}. To improve the Wasm runtimes' performance, Wasm runtime designers could pay attention to the optimization of WASI and other counterparts, such as EEI methods, by trading off the performance and tedious security and sandbox checks.

What's more, as for the \textbf{Wasm runtimes in various environments}, almost all the articles need to design a Wasm runtime completely and modify the existing runtimes to adapt to a new environment~\cite{chadha2023exploring, kolosick2022isolation, li2022bringing, peach2020ewasm, subramanyan2023privaton}. That is mainly due to the fixed WASI specification and Wasm runtime architecture. We advise decoupling different components in Wasm runtimes thoroughly so that researchers can combine the components they need to form a new Wasm runtime rather than modifying more than one part of an existing Wasm runtime or designing a new one~\cite{chadha2023exploring, kolosick2022isolation, li2022bringing, peach2020ewasm, subramanyan2023privaton}, such as design a standalone component to check the bounds of linear memory which could be applied to various Wasm runtimes as a hot plugging component.

\subsection{Wasm runtime testing}
Compared to the design and utilization of Wasm runtimes, testing for Wasm runtimes is not yet fully matured, thereby presenting numerous opportunities for improvement.

Although there are more than 20 articles focused on the performance testing of Wasm runtimes, there is still some space to improve, and we have some suggestions for \textbf{Wasm runtime performance testing} in the future:

1) \textit{Methodology.} The existing forms of performance testing for Wasm runtimes are relatively monolithic, typically involving the design of benchmarks followed by running them under different runtimes and configurations to compare execution times~\cite{oliveira2020analysis, sandhu2018sparse, yan2021understanding, mohan2022comparative, stotoglou2023comparative}. Only Kyriakou et al.~\cite{kyriakou2022complementing} considered stress testing. In the future, consider incorporating other methodologies that could help, such as concurrency testing, endurance testing, scalability testing, etc.

2) \textit{Benchmark.} Many benchmarks for testing are pretty simple, with few test cases, and may not be representative~\cite{oliveira2020analysis, aliyev2023analysis}. It might be beneficial to consider incorporating more representative Wasm benchmarks for performance testing in the future. For instance, the use of WasmBench~\cite{hilbig2021empirical}, containing more than 8,000 Wasm programs from the real world, could be considered.

3) \textit{High-level language.} The majority of Wasm test cases are compiled from C~\cite{mohan2022comparative, oliveira2020analysis, wang2021empowering, yan2021understanding, de2022webassembly}, and some are from Rust~\cite{jangda2019not} and Go~\cite{mohan2022comparative}. However, there is a lack of Wasm compiled from other high-level languages, such as Python and Java. The choice of languages may impact the execution performance of Wasm. Constructing datasets with equivalent functionality in different high-level languages is also challenging. Furthermore, while there are performance tests for Wasm compiled from various high-level languages, there is no fundamental analysis of why different runtimes may exhibit preferences for Wasm programs compiled from different high-level languages. Additionally, there is a lack of analysis regarding the impact of different high-level languages on the size of Wasm files. Currently, there is a lack of a comprehensive test evaluating the influence of high-level languages on Wasm binaries and Wasm runtimes.

4) \textit{Performance optimization.} While Jiang et al. have compared the performance of different runtimes, at most, it has only identified specific functions causing performance losses without analyzing the specific reasons or proposing optimization measures~\cite{jiang2023revealing}. In the future, optimizing the performance of Wasm runtimes is a promising approach.

5) \textit{Envrionment.} As Wasm's primary advantage of compatibility~\cite{Wasm-org}, it allows for varying CPU architectures(m), diverse OS platforms(n), and different runtimes(q), enabling the creation of $m \times n \times q$ types of testing platforms.

Currently, the \textbf{bug detection in Wasm runtimes} mainly uses black or grey box methods that could be applied to various runtimes~\cite{cao2023wrtester, zhou2023wadiff, jiang2022wasmfuzzer}. The bug location by Cao et al. only pinpoint which line of the Wasm instructions cause the bug, rather than finding which line of source code in Wasm runtimes cause the bug.
Introducing the white box method could detect and locate unknown bugs with domain knowledge in different Wasm runtimes, etc. Designing test generators for different components in the Wasm runtimes could help, such as proposing the test generator for WASI based on the WASI specification, which the current articles do not cover, etc.

\subsection{Directions for different stakeholers}
Besides designing or testing Wasm runtimes, there could be more directions to study for the stakeholders in the Wasm community.

For the \textbf{developers of the Wasm specification}, a more structured, decoupled Wasm specification would facilitate efficient development of Wasm runtimes. This would eliminate the need for extensive modifications at the source code level or create a runtime from scratch, promoting structured development and composition of Wasm runtimes. Additionally, as part of the Wasm specification, the imprecise description of the WASI specification has posed challenges to developing Wasm runtimes~\cite{zhang2023characterizing}. It would be better to clarify the WASI specification instead of relying solely on POSIX functions to represent the functionality of WASI functions, as is currently the case.

For \textbf{the Wasm runtime developers}, it is crucial to strictly adhere to the requirements of the Wasm specification in designing Wasm runtimes. Otherwise, executing the same program on different Wasm runtimes may yield different results, potentially causing issues such as calculation errors~\cite{cao2023wrtester}, inability to reach consensus in blockchain~\cite{zheng2020vm}, and so forth. In addition, ensuring the security of Wasm runtimes is also crucial, which is fundamental to why many users choose Wasm runtimes. Therefore, Wasm runtime developers need to ensure the integrity of the Wasm runtime sandbox mechanism and the security of linear memory, especially in multi-tenant scenarios. Wasm runtime developers should also strive to optimize the runtime's performance, which is particularly important in resource-constrained environments such as IoT and edge computing.

For \textbf{the testers of Wasm runtimes}, many aspects can be improved, such as methods, benchmarks, and so on. Testing of Wasm runtimes is still immature. Only by clearly and thoroughly testing and analyzing the issues in existing Wasm runtimes can we properly guide the design direction of future Wasm runtimes.

For \textbf{the users of Wasm runtimes}, they have already deployed Wasm runtimes in a wide range of scenarios, fully leveraging its lightweight, secure, and other features. Users can consider various factors when selecting the appropriate Wasm runtime for their needs, such as performance, compatibility with hardware devices in the scenario, and memory consumption of the runtime. Although some Wasm runtimes claim to be compatible with multiple platforms, the reality may differ. Users can refer to existing research results to choose the suitable platform rather than blindly trusting the official promotion of Wasm runtime.

\section{Related work}\label{sec:relatedwork}
There has been a large number of surveys focusing on different technology areas, such as serverless computing~\cite{wen2023rise, li2022serverless, eismann2021state}, machine learning~\cite{zhang2020machine, xie2023survey, baltruvsaitis2018multimodal, verbraeken2020survey}, software testing~\cite{barr2014oracle, garcia2023quantum, lima2023software}, etc. 
For example, Wen et al. presented a comprehensive overview of the existing literature to characterize the state of serverless computing~\cite{wen2023rise}. They collected and analyzed the 164 articles on 17 research directions about serverless computing and derived the research trends and focus. Zhang et al. presented a comprehensive survey of the methodologies used to test machine learning (ML) systems based on the collected 144 papers. Moreover, they discussed the trends of datasets and research hotspots and proposed promising future directions in ML testing~\cite{zhang2020machine}.
However, although Wasm runtime plays a significant role in the Wasm community, there is no survey about Wasm runtime research. We provide the first comprehensive survey of WSAM runtime research, following the methodologies used in the above surveys.

Currently, there are two surveys about Wasm, focusing on articles related to Wasm binaries~\cite{harnes2024sok, kim2022avengers}. Harnew et al. analyze techniques for Wasm binaries~\cite{harnes2024sok}, while Kim et al. analyze literature on methods used for studying Wasm binary security~\cite{kim2022avengers}. Our work presents the first comprehensive literature survey focusing on Wasm runtimes, aiming to contribute to the Wasm community.

\section{Threats to validity}\label{sec:discussion}
This comprehensive was conducted according to the established guidelines to alleviate potential threats to validity~\cite{zhang2020machine, hassan2021survey, lo2021systematic, wen2023rise}. However, there are still certain limitations, including the article searching and selection.

One primary threat is that the keywords used for paper searching may have limitations. For instance, some paper titles may not explicitly reflect any connection with Wasm but might involve Wasm runtimes. In our approach, there is a possibility of missing the collection of these articles. However, to gather papers as comprehensively as possible, the search keywords were jointly proposed by the first three authors, all of whom have over three years of research experience in the field of Wasm.

Another primary threat is the selection of articles. To filter out papers related to Wasm runtime from the initially collected 498 articles, we excluded a portion of them, which may have led to some papers being erroneously filtered out. To minimize this possibility and retain papers closely related to Wasm runtime as much as possible, we established strict criteria for paper selection. We only filtered out articles that were duplicates, unavailable or did not analyze or modify Wasm runtime in any way.

\section{Conclusion}\label{sec:conclusion}
We provide a comprehensive outline and deeply analyze the research articles about Wasm runtimes. 
We first presented the definition, architecture, and current article state of Wasm runtimes. Second, we illustrated how the researchers designed a Wasm runtime entirely or added features based on existing Wasm runtimes. Third, we presented the benchmarks, methods, and results of testing Wasm runtimes and illustrated how researchers detect bugs in Wasm runtimes. Finally, we discussed the problems in current articles and proposed future research directions for Wasm runtimes and the whole Wasm community. 
We hope this survey can help Wasm runtime researchers, developers, users, and the developers of Wasm applications become familiar with the current development state and simultaneously provide research opportunities.

\bibliographystyle{ACM-Reference-Format}
\bibliography{sample-base}

\end{document}